\documentclass[aps]{revtex4}
\usepackage[utf8]{inputenc}
\usepackage{amsfonts}
\usepackage{amsmath}
\usepackage{amssymb,epsf}

\begin{document}
\title{Black Holes in Gauss-Bonnet Gravity's Rainbow}
\author{Seyed Hossein Hendi$^{1,2}$\footnote{email address: hendi@shirazu.ac.ir} and Mir Faizal$^3$ \footnote{email address: f2mir@uwaterloo.ca}}
\affiliation{$^1$ Physics Department and Biruni Observatory,
College of Sciences, Shiraz University, Shiraz 71454, Iran\\
$^2$ Research Institute for Astronomy and Astrophysics of Maragha
(RIAAM), Maragha, Iran\\
$^3$ Department of Physics and Astronomy, University of Waterloo,
Waterloo, Ontario, N2L 3G1, Canada}
\date{29-June-2015}

\begin{abstract}
In this paper, we will generalize the Gauss-Bonnet gravity to an
energy dependent Gauss-Bonnet theory of gravity, which we shall
call as the Gauss-Bonnet gravity's rainbow. We will also couple
this theory to a Maxwell's theory. We will analyze black hole
solutions in this energy dependent Gauss-Bonnet gravity's rainbow.
We will calculate the modifications to the thermodynamics of black
holes in the Gauss-Bonnet's gravity's rainbow. We will demonstrate
that even though the thermodynamics of the black holes get
modified in the Gauss-Bonnet gravity's rainbow, the first law of
thermodynamics still holds for this modified thermodynamics. We
will also comment on the thermal stability of the black hole
solutions in this theory.
\end{abstract}

\maketitle

\section{Introduction\label{Intro}}

The form of the standard energy-momentum dispersion relation is
fixed by the Lorentz symmetry. Even though the Lorentz symmetry is
one of the most important symmetries in nature, there are
indications from various approaches to quantum gravity that the
Lorentz symmetry might be violated in the ultraviolet limit
\cite{l}-\cite{l1}. Thus, it is possible that the Lorentz symmetry
is only an effective symmetry which holds in the infrared limit of
quantum gravitational processes. As the standard energy-momentum
dispersion relation depends on the Lorentz symmetry, it is
expected that the standard energy-momentum dispersion relation
will also get modified in the ultraviolet limit. In fact, it has
been observed that such modification to the standard
energy-momentum relation does occur in the discrete spacetime
\cite{1}, models based on string field theory \cite{2}, spacetime
foam \cite{4}, the spin-network in loop quantum gravity (LQG)
\cite{5}, non-commutative geometry \cite{6}, and Horava-Lifshitz
gravity \cite{7}-\cite{7a}.

The modification of the  standard energy-momentum dispersion relation has
motivated the development of double special relativity \cite{8}. In this
theory, apart from the velocity of light being the maximum velocity
attainable, there is also a maximum energy scale in nature. This energy
scale is the Planck energy $E_P$, and it is not possible to for a particle
to attain energies beyond this energy. The double special relativity has
been generalized to curved spacetime, and this doubly general theory of
relativity is called gravity's rainbow \cite{9}. In this theory, the
geometry of spacetime depends on the energy of the test particle. So,
particles of different energy see the geometry of spacetime differently.
Hence, the geometry of spacetime is represented by a family of energy
dependent metrics forming a rainbow of metrics. This is the reason the
theory has been called gravity's rainbow. In order to construct this theory,
the modified energy-momentum dispersion relation is written as
\begin{equation}  \label{MDR}
E^2f^2(E/E_P)-p^2g^2(E/E_P)=m^2
\end{equation}
where $E_P$ is the Planck energy. The functions $f(E/E_P)$ and $g(E/E_P)$
are called rainbow functions, and they are required to satisfy
\begin{equation}
\lim\limits_{E/E_P\to0} f(E/E_P)=1,\qquad \lim\limits_{E/E_P\to0} g(E/E_P)=1.
\end{equation}
This condition is needed, as the theory is constrained to
reproduce the standard dispersion relation in the infrared limit.
Motivated by this energy dependent modification to the dispersion
relation, the metric $h(E)$ in gravity's rainbow is written as
\cite{10}
\begin{equation}  \label{rainmetric}
h(E)=\eta^{ab}e_a(E)\otimes e_b(E).
\end{equation}
Here the energy dependence of the frame fields can be written as
\begin{equation}
e_0(E)=\frac{1}{f(E/E_P)}\tilde{e}_0, \qquad e_i(E)=\frac{1}{g(E/E_P)}\tilde{%
e}_i,
\end{equation}
where the tilde quantities refer to the energy independent frame
fields. The energy at which the spacetime is probed is represented
by $E$, and the maximum attainable energy is represented by $E_p$.
So, if a text particle is used to probe the geometry of spacetime,
then $E$ is the energy of that test particle. Thus, by definition
$E$ cannot become greater than $E_p$ \cite{10}. As we will be
applying gravity's rainbow to the black hole thermodynamics, the
energy $E$ will correspond to the energy a quantum particle in the
vicinity of the event horizon, which is emitted in the Hawking
radiation \cite{5A}-\cite{5Q}. It is possible to translate the
uncertainty principle $\Delta p \geq 1/\Delta x $ into a bound on
energy $E \geq 1/\Delta x $ \cite{5T}. In gravity's rainbow, even
though the metric depends on the energy, the  usual uncertainty
principle still holds \cite{1A}. It has been demonstrated that the
uncertainty in the position of the particle near the horizon
should be equal to the radius of the event horizon radius
\cite{5A}-\cite{5Q}
\begin{equation}
 E\geq 1/{\Delta x} \approx 1/{r_+}.
\end{equation}
This energy $E$ of the particle in the vicinity of the event
horizon will be used in the rainbow functions. It may be noted
that this energy $E$ is a dynamical function of the radial
coordinate \cite{5D}. We will not need the explicit dependence of
this energy on the radial coordinate, but it is important to note
that the rainbow functions are dynamical, and so they cannot be
gauged away. Furthermore, this energy of a particle near the
horizon is bounded by the Planck energy $E_p$, and it cannot
increase to an arbitrary value.

This bound on the energy also modifies the temperature of the
black, and this modification to the temperature of the black hole
can be used to calculate the corrections to the entropy of the
black hole in gravity's rainbow \cite{5A}. The modification of the
thermodynamics of black rings in gravity's rainbow has also been
calculated \cite{2A}. It has been observed that the temperature of
both the black holes and black rings starts to reduce after
attaining a maximum value. Thus, at a critical size the
temperature of the black holes and black rings becomes zero. At
this value the entropy also becomes zero. Thus, the back hole
stops radiating Hawking radiation, after reaching this critical
radius. Thus, the gravity's rainbow predicts the existence of a
black remnant. It has been argued that such a remnant will form
for all black objects \cite{4A}. In fact, it has been explicitly
demonstrated that such a remnant does form for Kerr black holes,
Kerr-Newman black holes in de Sitter space, charged AdS black
holes, higher dimensional Kerr-AdS black holes and black saturn
\cite{4A}.

It may be noted that the behavior of gravity's rainbow depends on
the form of the rainbow functions chosen. The construction of
these rainbow functions is motivated from various theoretical and
phenomenological considerations. Even though the form of these
rainbow functions is different, the main property of these rainbow
functions is that they make the geometry of spacetime dependent on
the energy of the probe. This is the property which we will
require in our analysis. So, in  this paper, we will analyze the
behavior of black hole solutions in the  Gauss-Bonnet gravity's
rainbow using these phenomenologically motivated rainbow
functions.

\section{ Gauss-Bonnet Gravity's Rainbow}

It may be noted that it is possible to add higher order curvature terms to
the original Lagrangian for general relativity. The theory obtained by
adding quadratic powers of the curvature terms is called Gauss-Bonnet
gravity. It is also possible to construct a Gauss-Bonnet-Maxwell gravity
\cite{gb}-\cite{bg}. The Lagrangian of Gauss-Bonnet-Maxwell gravity can be
written as
\begin{equation}
L_{\mathrm{tot}}=R-2\Lambda +\alpha (R_{abcd}R^{abcd}-4R_{ab}R^{ab}+R^{2})-%
\mathcal{F},  \label{Lagrangian}
\end{equation}%
where $R_{abcd}$ and $\Lambda $ are the Riemann tensor and the
cosmological constant, respectively. Here $\alpha $ denotes the
Gauss-Bonnet (GB) coefficient, $R= R_{a}^{a}$ denotes the Ricci
scalar  and $R_{cd}=R_{cad}^{a} $ denotes the Ricci tensor. The
last term in Eq. (\ref{Lagrangian}) is the Maxwell invariant
$\mathcal{F}=F_{ab}F^{ab}$, where $F_{ab}=\partial
_{a}A_{b}-\partial _{b}A_{a}$ is the electromagnetic field tensor
and $A_{b}$ is the gauge potential. Now the fields equations for
the the Gauss-Bonnet-Maxwell gravity Lagrangian (\ref{Lagrangian})
can be written as
\begin{equation}
G_{ab}^{E}+\Lambda g_{ab}+\alpha G_{ab}^{GB}=-\frac{1}{2}g_{ab}\mathcal{F}%
+2F_{ac}F_{b}^{c},  \label{Feq1}
\end{equation}%
where
\begin{equation}
\nabla _{a}F^{ab}=0.  \label{Feq2}
\end{equation}%
Here the the Einstein tensor is denoted by $G_{ab}^{E}$ and
\begin{eqnarray*}
G_{ab}^{GB} &=&2\left(
R_{acde}R_{b}^{cde}-2R_{acbd}R^{cd}-2R_{ac}R_{b}^{c}+RR_{ab}\right) - \\
&&\frac{1}{2}\left( R_{cdef}R^{cdef}-4R_{cd}R^{cd}+R^{2}\right) g_{ab}.
\end{eqnarray*}
In gravity's rainbow the spacetime geometry depends on the energy of the
probe ($E$). We can follow the methods used in the usual gravity's rainbow
\cite{8}, and absorb all the energy dependence of the Gauss-Bonnet gravity's
rainbow into the rainbow functions. So, we can write the  energy dependent
metric Gauss-Bonnet gravity's rainbow as
\begin{equation}
d\tau ^{2}=-ds^{2}=\frac{\Psi (r)}{f(E)^{2}}dt^{2}-\frac{1}{g(E)^{2}}\left(
\frac{dr^{2}}{\Psi (r)}+{r^{2}}d\Omega ^{2}\right)  \label{Metric}
\end{equation}
where
\begin{equation}
d\Omega ^{2}=d\theta
_{1}^{2}+\sum\limits_{i=2}^{d-2}\prod\limits_{j=1}^{i-1}\sin
^{2}\theta _{j}d\theta _{i}^{2}  \label{dOmega2}
\end{equation}
Since we are looking for the black hole solutions with a radial
electric field, we know that the nonzero components of the
electromagnetic field are
\begin{equation}
F_{tr}=-F_{rt}.  \label{nonzero}
\end{equation}
One can use Eq. (\ref{Feq2}) with the metric (\ref{Metric}) to
obtain the following explicit form of $F_{tr}$
\begin{equation}
F_{tr}=\frac{q}{r^{d-2}},  \label{Ftr}
\end{equation}
where $q$ is an integration constant related to the electric
charge of the black hole. In addition, it is easy to show that the
following metric function satisfies all of the field equations
(\ref{Feq1}), simultaneously
\begin{equation}
\Psi (r)=1+\frac{r^{2}}{2\alpha ^{\prime }g(E)^{2}}\left( 1-\sqrt{\Theta (r)}%
\right) ,  \label{Psi}
\end{equation}%
with
\begin{equation}
\Theta (r)= 1+\frac{8\alpha ^{\prime }}{(d-1)(d-2)}\left( \Lambda
+\frac{(d-1)(d-2)m}{2r^{d-1}}-\frac{(d-1)(d-3)q^{2}f(E)^{2}g(E)^{2}}{r^{2d-4}}\right),
\label{Th}
\end{equation}%
where $m$ is an integration constant that is related to mass and
$\alpha ^{\prime }=(d-3)(d-4)\alpha$. Taking into account Eqs.
(\ref{Ftr}) and (\ref{Psi}), it is clear that one cannot remove
rainbow functions with rescaling.

The choice of the rainbow functions $f(E/E_{p})$ and $g(E/E_{p})$
is very important for making predictions. That choice is preferred
to be based on phenomenological motivations. Many proposals exist
in the literature, we will some forms which have important
phenomenological motivations.

The rainbow functions motivated by the results obtained from loop quantum
gravity and non-commutative geometry are given by \cite{1z}-\cite{z1}
\begin{equation}
f\left( E/{{E}_{p}}\right) =1,\quad g\left( E/{{E}_{p}}\right) =\sqrt{1-\eta
\left( E/E_{p}\right) ^{n}}.  \label{Mod1}
\end{equation}%
Now using these rainbow functions the metric is given by
\begin{equation}
\Psi _{1}(r)=1+\frac{r^{2}}{2\alpha ^{\prime }\left[ 1-\eta \left( \frac{E}{%
E_{p}}\right) ^{n}\right] }\left( 1-\sqrt{\Theta _{1}(r)}\right) ,
\label{Psi1}
\end{equation}%
with
\begin{equation}
\Theta _{1}(r)=1+\frac{8\alpha ^{\prime }}{(d-1)(d-2)}\left( \Lambda +%
\frac{(d-1)(d-2)m}{2r^{d-1}}-\frac{(d-1)(d-3)q^{2}\left[ 1-\eta
\left( \frac{E}{E_{p}}\right) ^{n}\right] }{r^{2d-4}}\right) .
\label{Th1}
\end{equation}

The hard spectra from gamma-ray burster's are given by can also be used to
motivate the construction of rainbow functions \cite{2z}. These rainbow
functions are given by
\begin{equation}
f\left( E/{{E}_{p}}\right) =\frac{{{e}^{\beta E/Ep}}-1}{\beta E/{{E}_{p}}}%
,\quad g\left( E/{{E}_{p}}\right) =1.  \label{Mod3}
\end{equation}%
The metric corresponding to Eqs. (\ref{Psi}), (\ref{Th}) and (\ref{Mod3}),
can be written as
\begin{equation}
\Psi _{2}(r)=1+\frac{r^{2}}{2\alpha ^{\prime }}\left( 1-\sqrt{\Theta _{2}(r)}%
\right) ,  \label{Psi3}
\end{equation}%
with
\begin{equation}
\Theta _{2}(r)=1+\frac{8\alpha ^{\prime }}{(d-1)(d-2)}\left( \Lambda +%
\frac{(d-1)(d-2)m}{2r^{d-1}} -\frac{(d-1)(d-3)q^{2}{{E}_{p}^{2}}\left( {{e}^{\frac{\beta E}{Ep}}%
}-1\right) ^{2}}{\beta ^{2}E^{2}r^{2d-4}}\right) ,  \label{Th3}
\end{equation}
rainbow functions in which the velocity of light is constant have
also been analyzed. This choice of rainbow functions is given by
\cite{4z}
\begin{equation}
f\left( E/{{E}_{p}}\right) =g\left( E/{{E}_{p}}\right) =\frac{1}{1-\lambda E/%
{{E}_{p}}}.  \label{Mod4}
\end{equation}
The metric function corresponding to the case where the velocity
of light is a constant can be written as (\ref{Mod3})
\begin{equation}
\Psi_{3}(r)=1+\frac{r^{2}\left( 1-\frac{\lambda E}{{{E}_{p}}}\right) ^{2}}{%
2\alpha ^{\prime }}\left( 1-\sqrt{\Theta _{3}(r)}\right) ,  \label{Psi4}
\end{equation}
with
\begin{equation}
\Theta _{3}(r) = 1+\frac{8\alpha ^{\prime }}{(d-1)(d-2)}\left( \Lambda +%
\frac{(d-1)(d-2)m}{2r^{d-1}}-\frac{(d-1)(d-3)q^{2}}{r^{2d-4}\left( 1-\frac{\lambda E}{{{E}_{p}}}%
\right) ^{4}}\right).  \label{Th4}
\end{equation}
It may be noted that the bounds on the values of $\beta$, $\eta$,
$\lambda$ has been analyzed using various theoretical and
experimental considerations \cite{5z}.

\section{Thermodynamics and Thermal Stability \label{Asympflat}}

In this section, we will analyze the thermodynamics of the black
hole solution obtained in the previous section. It is possible to
obtain the Hawking temperature by using the surface gravity,
\begin{equation}
T=\frac{g(E)}{r_{+}f(E)}\frac{(d-2)(d-3)\left( 1+\frac{\alpha ^{\prime
}(d-5)g^{2}(E)}{(d-3)r_{+}^{2}}\right) -\frac{2\Lambda r_{+}^{2}}{g^{2}(E)}-%
\frac{2f^{2}(E)(d-3)^{2}q^{2}}{r_{+}^{2d-6}}}{4\pi (d-2)\left( 1+\frac{%
2\alpha ^{\prime }g^{2}(E)}{r_{+}^{2}}\right) }.  \label{T}
\end{equation}

According to the area law, the entropy of black holes is equal to
one-quarter of the horizon area. This relation is valid for
Einstein gravity, whereas we are not allowed to use it for higher
derivative gravity. In this paper we investigate it in
Gauss-Bonnet gravity's rainbow. We can use the Wald formula for
calculating the entropy
\begin{equation}
S=\frac{1}{4}\int d^{n-1}x\sqrt{\gamma }\left( 1+2\alpha \widetilde{R}\right)
\label{Swald1}
\end{equation}%
where $\widetilde{R}$ is the Ricci scalar for the induced metric
$\gamma _{ab}$ on the $\left( d-2\right) $ dimensional boundary.
We obtain
\begin{equation}
S=\frac{V_{d-2}r_{+}^{d-2}}{4g^{d-2}(E)}\left( 1+\frac{2\left( d-2\right)
\alpha ^{\prime }g^{2}(E)}{(d-4)r_{+}^{2}}\right) ,  \label{S}
\end{equation}%
which confirms that the obtained black hole solutions violate the
area law. We should mention that $V_{d-2}$ denotes the volume of
$(d-2)$-dimensional sphere.

Considering the Gauss law and calculating the flux of the electric
field at infinity, one can find the electric charge of the black
hole has the following form
\begin{equation}
Q=\frac{V_{d-2}(d-3)q}{4\pi }\frac{f(E)}{g^{d-3}(E)}.  \label{Q}
\end{equation}
We can calculate the electric potential $\Phi $ as
\begin{equation}
\Phi =\left. A_{\mu }\chi ^{\mu }\right\vert _{r\longrightarrow \infty
}-\left. A_{\mu }\chi ^{\mu }\right\vert _{r\longrightarrow r_{+}}=\frac{q}{%
r_{+}^{d-3}},  \label{Phi}
\end{equation}%
where $\chi ^{\mu }$ is the null generator of the horizon.

In order to calculate the finite mass of the black hole we will use the ADM
(Arnowitt-Deser-Misner) approach for large values of $r$
\begin{equation}
M=\frac{V_{d-2}}{16\pi }\frac{m\left( d-2\right) }{f(E)g^{d-1}(E)}.
\label{Mass}
\end{equation}
Now using Eqs. (\ref{Psi}), (\ref{S}) and (\ref{Q}) and
considering $M$ as a function of the extensive parameters $S$ and
$Q$, we obtain the following result,
\begin{eqnarray}
M\left( S,Q\right) &=& \frac{\left( d-2\right) }{16\pi f(E)g^{d-1}(E)}\left[
g^{2}(E)r_{+}^{d-3}\left( 1+\frac{\alpha ^{\prime }g^{2}(E)}{r_{+}^{2}}%
\right)\right.  \nonumber \\
&& \left. -\frac{2\Lambda r_{+}^{d-1}}{(d-1)(d-2)} +\frac{32\pi
^{2}Q^{2}g^{2d-4}(E)}{(d-2)(d-3)r_{+}^{d-3}}\right] .  \label{Msmar}
\end{eqnarray}

We can calculate the temperature and electric potential as the intensive
parameter using the following relation
\begin{equation}
T=\left( \frac{\partial M}{\partial S}\right) _{Q}=\left( \frac{\partial M}{%
\partial r_{+}}\right) _{Q}\left/ \left( \frac{\partial S}{\partial r_{+}}%
\right) _{Q}\right. \ \ ,\ \ \Phi =\left( \frac{\partial M}{\partial Q}%
\right) _{S}.\   \label{TPhi}
\end{equation}%
It may be noted that these relations are similar to the relations obtained
in Eqs. (\ref{T}) and (\ref{Phi}. Thus, the conserved and thermodynamic
quantities satisfy the first law of thermodynamics,
\begin{equation}
dM=TdS+\Phi dQ.  \label{Firstk1}
\end{equation}

It may be noted that we can investigate the thermal stability of
the charged black hole solutions of Gauss-Bonnet gravity's rainbow
through the canonical ensemble. In this ensemble, the electric
charge is set as a fixed parameter and, therefore, the positivity
of the heat capacity
\begin{equation}
C_{Q}=\frac{T_{+}}{\left( \frac{\partial ^{2}M}{\partial
S^{2}}\right) _{Q}},
\end{equation}%
is sufficient to ensure the local stability. Since we are
investigating the physical black hole solutions with positive
temperature, it is sufficient to examine the positivity of
\begin{equation}
\left( \frac{\partial ^{2}M}{\partial S^{2}}\right) _{Q}=-\frac{%
(d-2)r_{+}^{d-4}\left[ (d-4)\alpha
g^{2}(E)+\frac{r_{+}^{2}}{2(d-3)}\right] ^{2}\Upsilon
_{1}}{2g^{d-2}(E)\Upsilon _{2}},  \label{dMdSSk1}
\end{equation}%
where
\begin{eqnarray}
\Upsilon _{1} &=&-\frac{2(d-3)q^{2}f^{2}(E)g^{2}(E)}{(d-2)r_{+}^{2d-8}}%
+(d-4)(d-5)\alpha g^{4}(E)+g^{2}(E)r_{+}^{2}  \nonumber \\
&&-\frac{2\Lambda r_{+}^{4}}{(d-2)(d-3)},
\end{eqnarray}
\begin{eqnarray}
\Upsilon _{2} &=&-\frac{2(d-3)q^{2}f^{2}(E)g^{2}(E)\left[
(d-4)(2d-7)\alpha g^{2}(E)+\frac{(2d-5)r_{+}^{2}}{2(d-3)}\right]
}{(d-2)r_{+}^{2d-8}}+   \nonumber \\
&&(d-4)^{2}(d-5)\alpha ^{2}g^{6}(E)+\frac{(d-4)(d-9)\alpha
g^{4}(E)r_{+}^{2}}{2(d-3)}+ \nonumber\\
&&\frac{g^{2}(E)\left[ d-2+12\alpha \Lambda \left( d-4\right)
\right] r_{+}^{4}}{2(d-2)(d-3)}+\frac{\Lambda
r_{+}^{6}}{(d-2)(d-3)^{2}}.
\end{eqnarray}
Thus, we are able to derive an explicit expression for the energy
dependence of the heat capacity of a black hole in Gauss-Bonnet
gravity's rainbow. The positivity of this heat capacity ensures
the local stability.

\begin{figure}[tbp]
$%
\begin{array}{cc}
\epsfxsize=6cm \epsffile{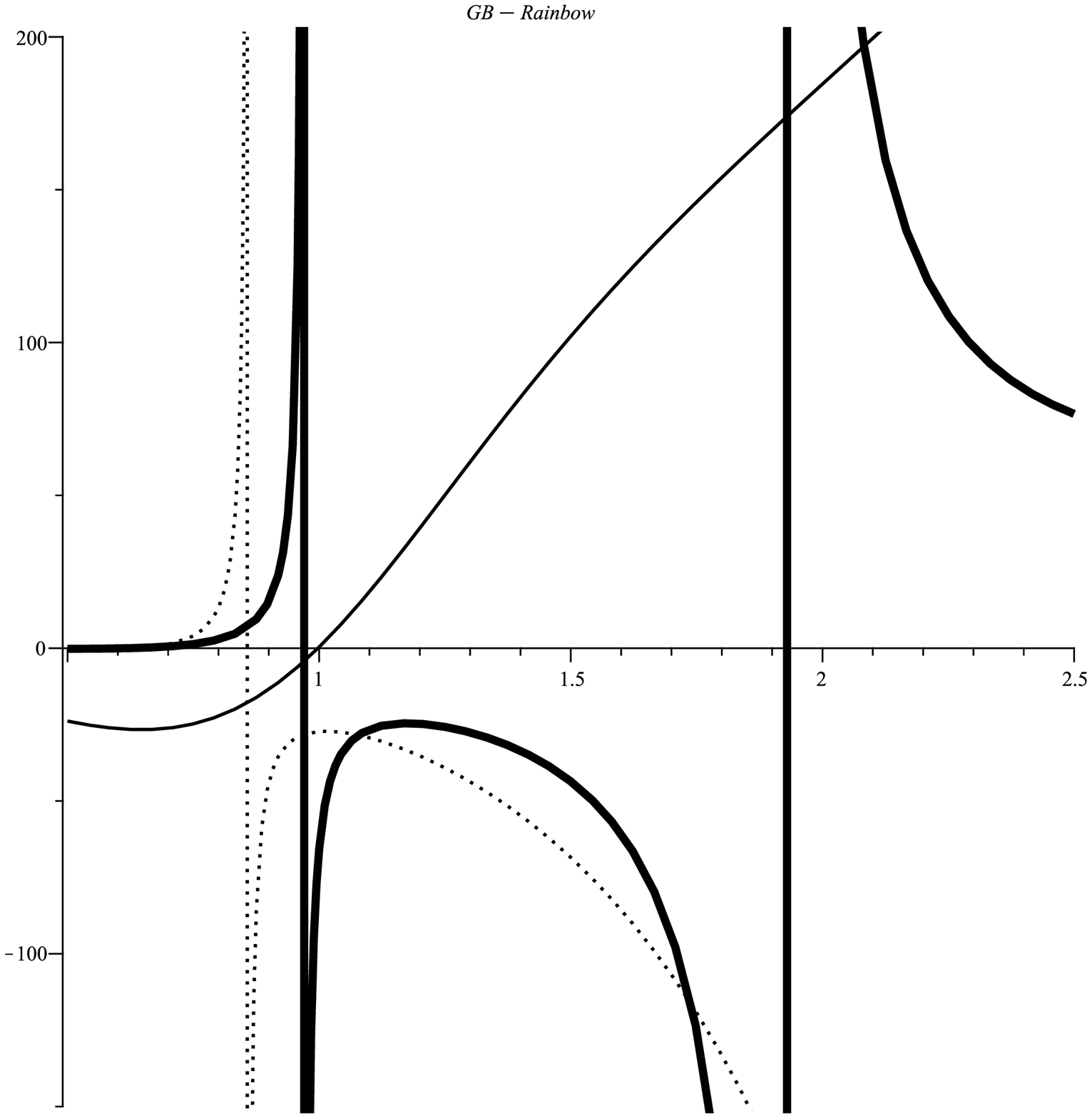} & \epsfxsize=6cm %
\epsffile{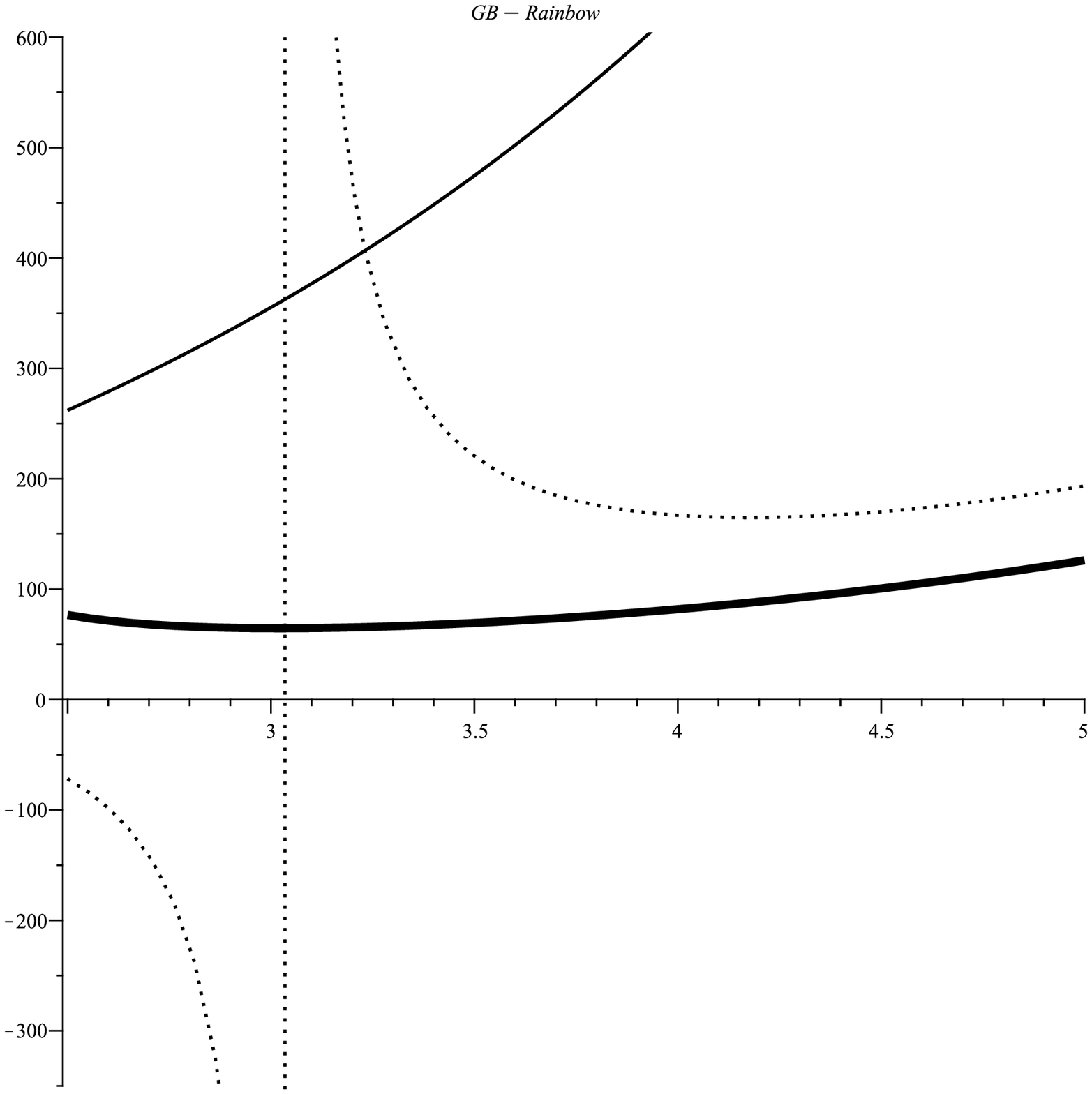}%
\end{array}
$.
\caption{\textbf{Model 1:} $C_{Q}$ versus $r_{+}$ for $E=1$, $E_{p}=5$, $\Lambda =-1$, $q=1$, $%
\protect\eta=1$, $n=2$, $\protect\alpha=5$, and $d=5$ (solid
line), $d=6$ (bold line) and $d=7$ (dotted line). \emph{"different
scales"}} \label{FigCdimensionM1}
\end{figure}
\begin{figure}[tbp]
$%
\begin{array}{cc}
\epsfxsize=6cm \epsffile{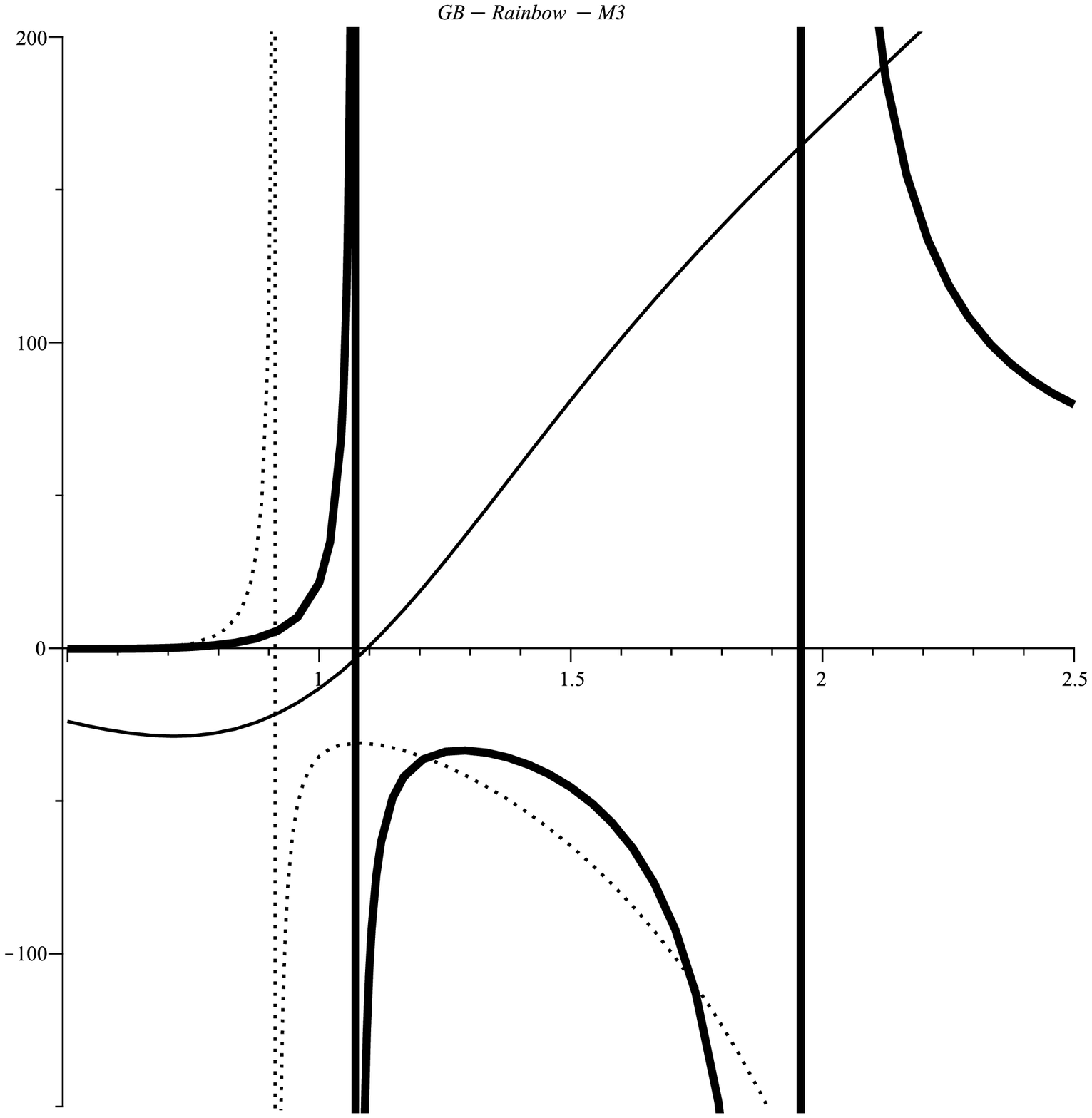} & \epsfxsize=6cm %
\epsffile{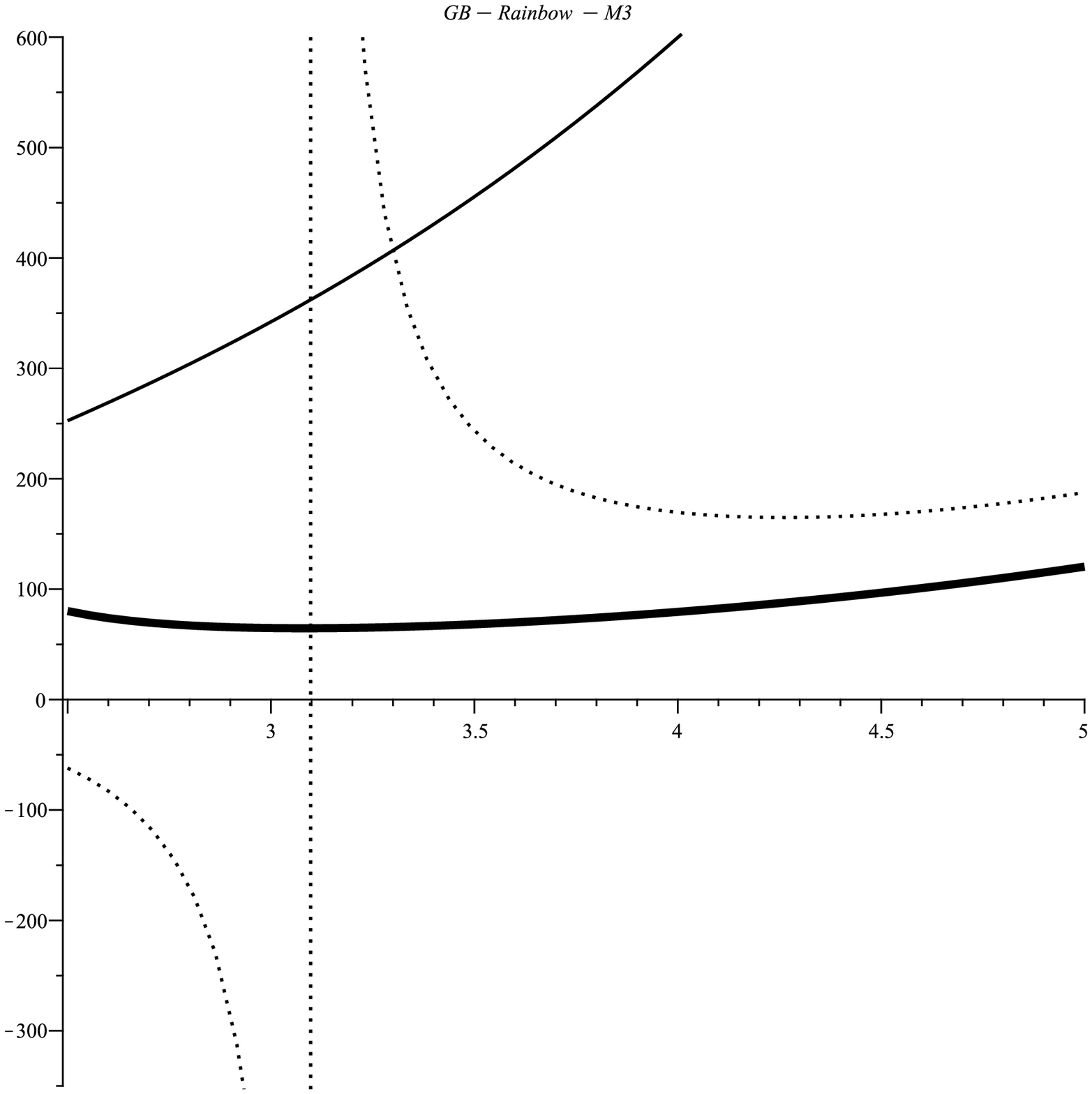}%
\end{array}
$. \caption{\textbf{Model 2:} $C_{Q}$ versus $r_{+}$ for $E=1$,
$E_{p}=5$, $\Lambda =-1$, $q=1$, $\protect\beta=2$,
$\protect\alpha=5$, and $d=5$ (solid line), $d=6$ (bold line) and
$d=7$ (dotted line). \emph{"different scales"}}
\label{FigCdimensionM3}
\end{figure}
\begin{figure}[tbp]
$%
\begin{array}{cc}
\epsfxsize=6cm \epsffile{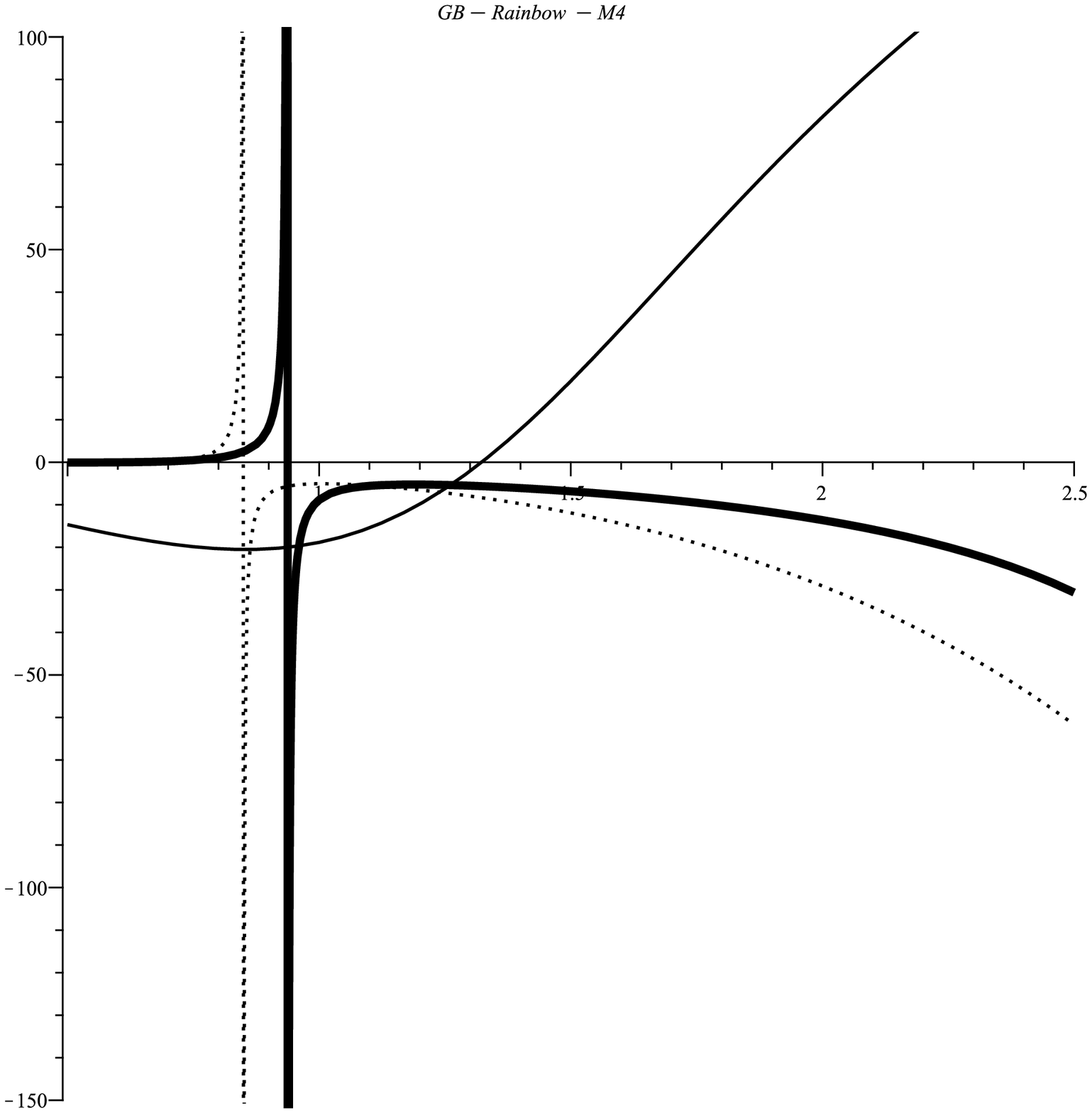} & \epsfxsize=6cm
\epsffile{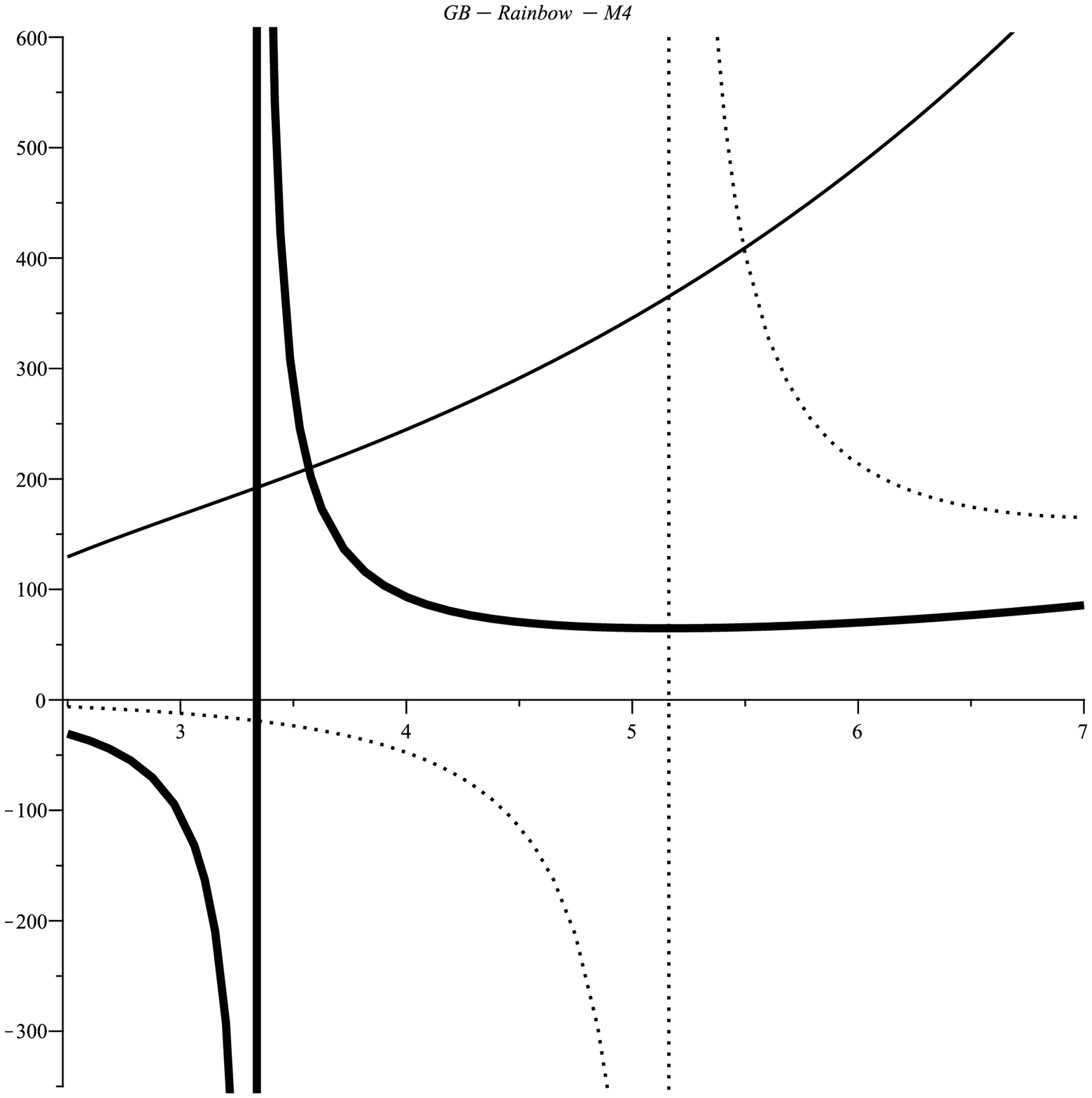}
\end{array}
$.
\caption{\textbf{Model 3:} $C_{Q}$ versus $r_{+}$ for $E=1$, $E_{p}=5$, $\Lambda =-1$, $q=1$, $%
\protect\lambda=2$, $\protect\alpha=5$, and $d=5$ (solid line),
$d=6$ (bold line) and $d=7$ (dotted line). \emph{"different
scales"}} \label{FigCdimensionM4}
\end{figure}

\begin{figure}[tbp]
$%
\begin{array}{cc}
\epsfxsize=6cm \epsffile{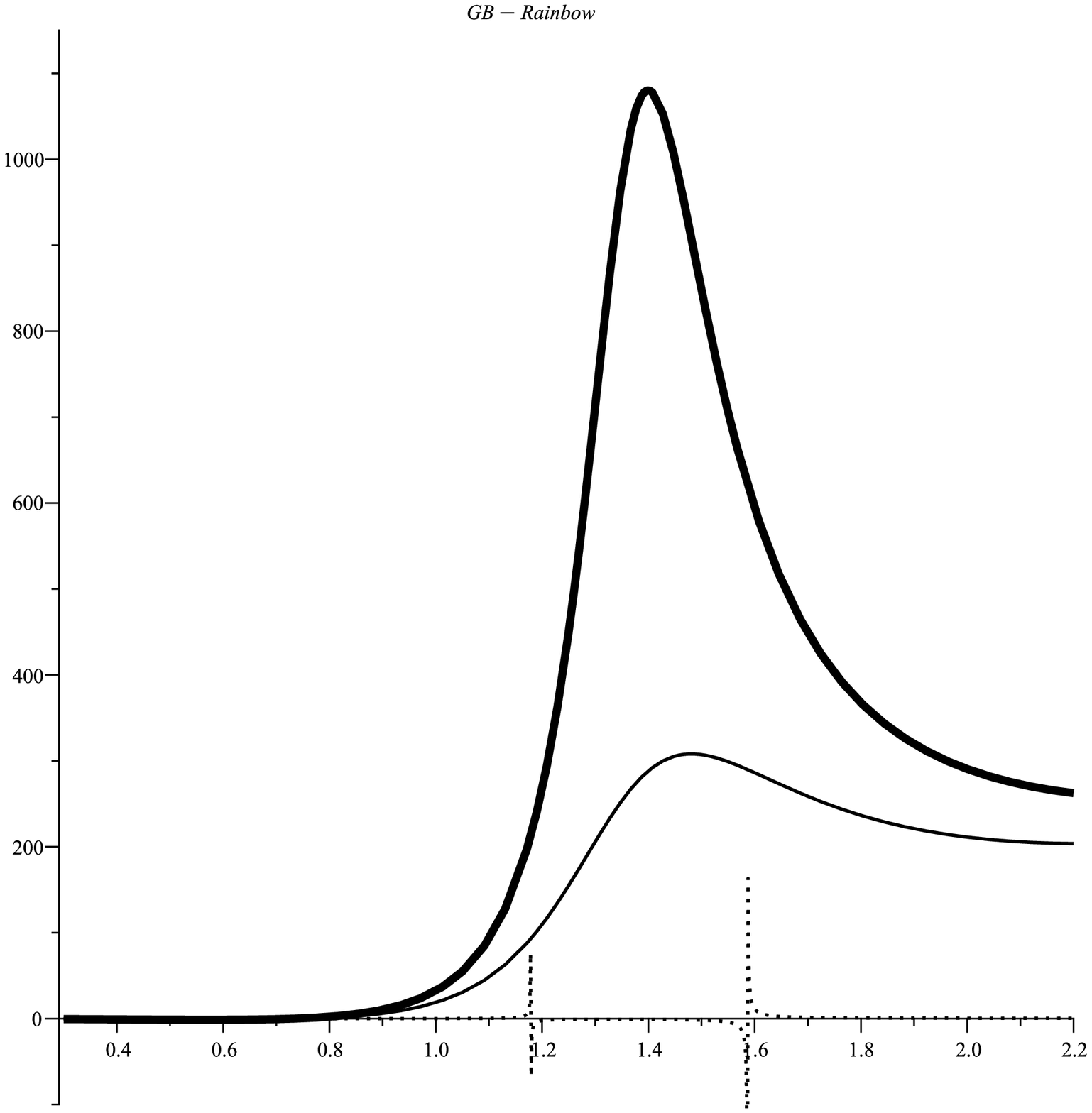} & \epsfxsize=6cm \epsffile{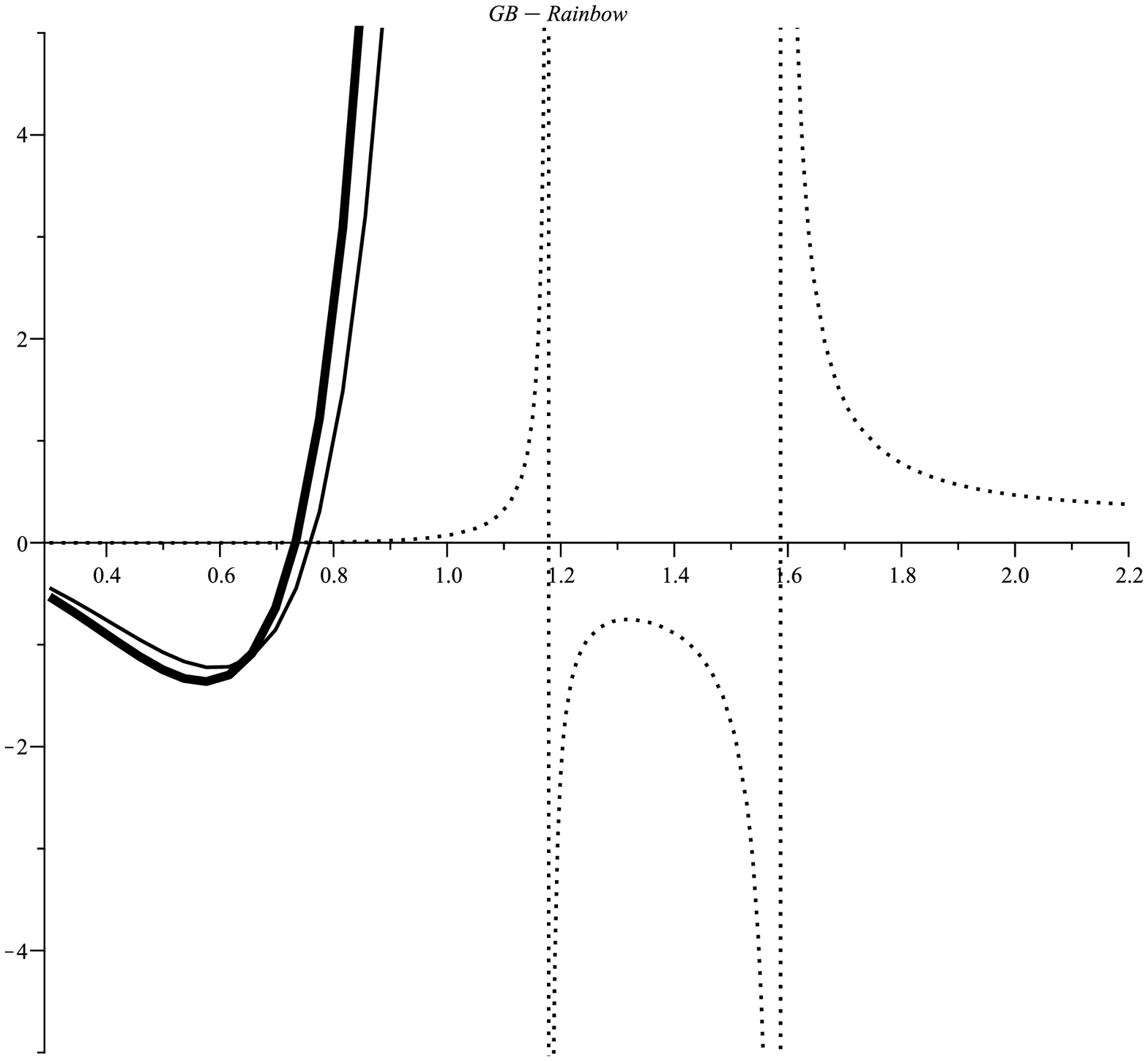}%
\end{array}
$.
\caption{\textbf{Model 1:} $C_{Q}$ versus $r_{+}$ for $E=1$, $E_{p}=5$, $\Lambda =-1$, $q=1$, $%
\protect\eta=1$, $n=2$, $d=6$, and $\protect\alpha=2$ (solid line), $\protect%
\alpha=2.4$ (bold line) and $\protect\alpha=3$ (dotted line). \emph{%
"different scales"}} \label{FigCalphaM1}
\end{figure}
\begin{figure}[tbp]
$%
\begin{array}{cc}
\epsfxsize=6cm \epsffile{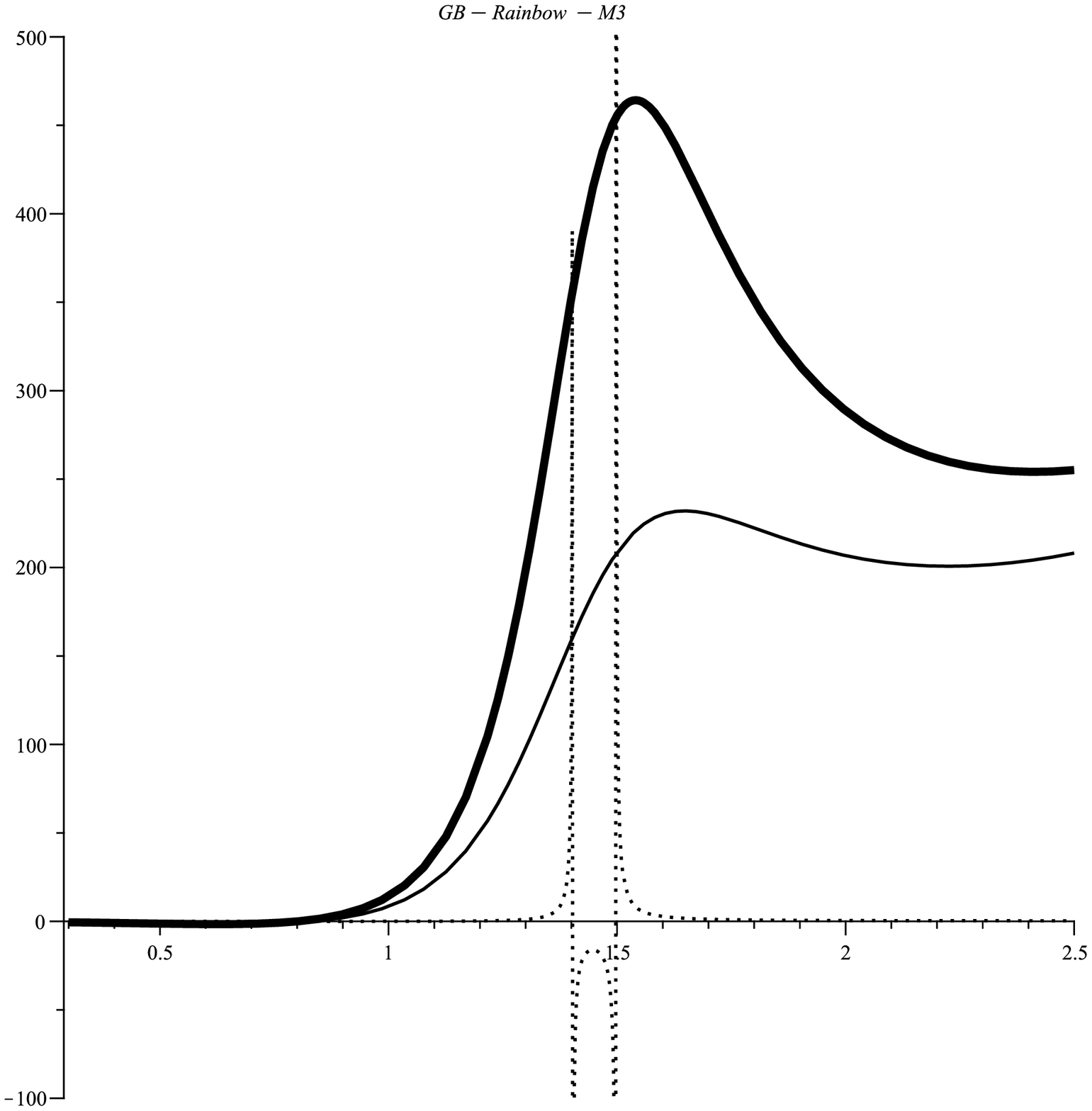} & \epsfxsize=6cm \epsffile{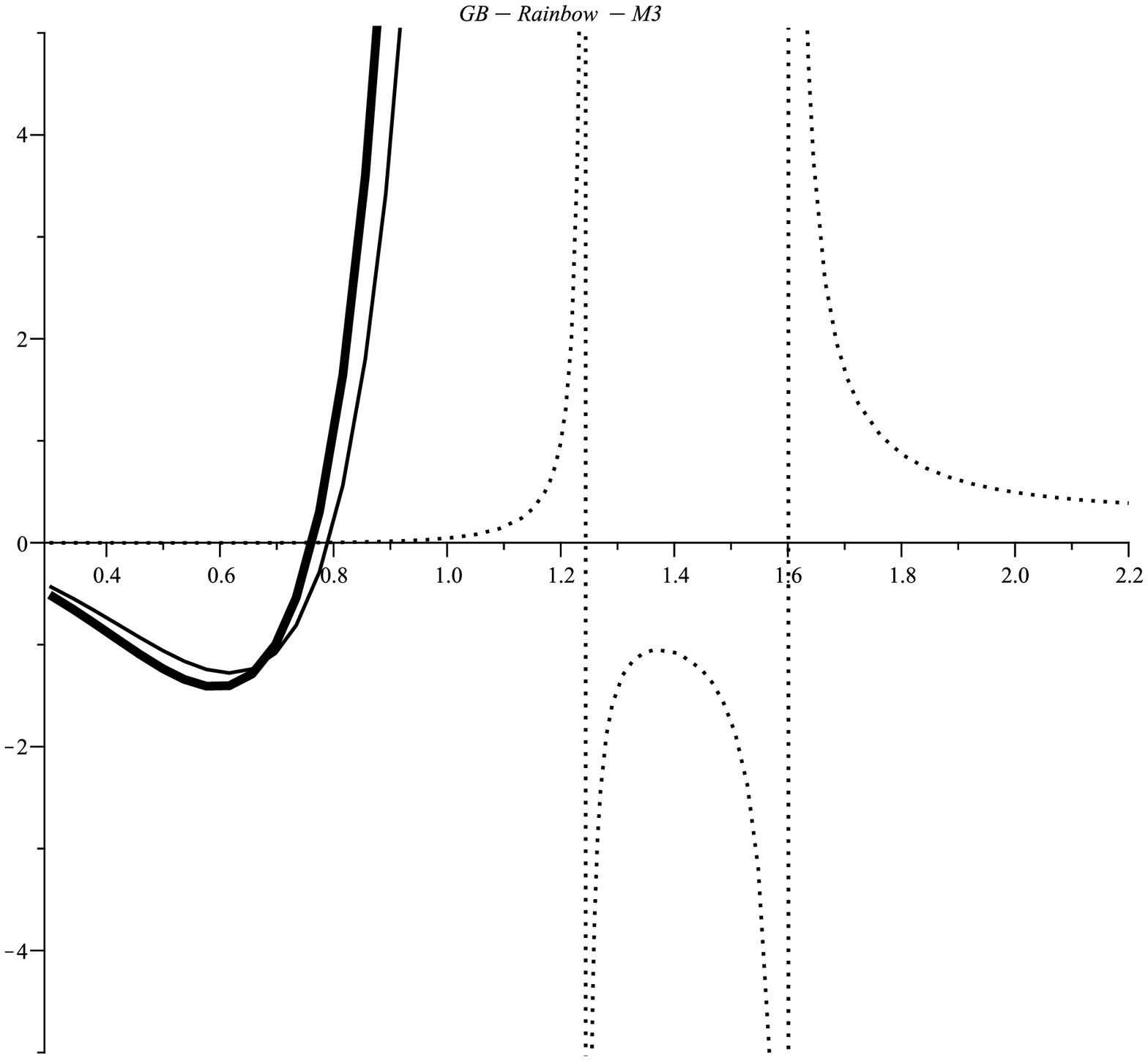}%
\end{array}
$. \caption{\textbf{Model 2:} $C_{Q}$ versus $r_{+}$ for $E=1$,
$E_{p}=5$, $\Lambda =-1$, $q=1$, $\protect\beta=2$, $d=6$, and
$\protect\alpha=2$ (solid line), $\protect\alpha=2.4$ (bold line)
and $\protect\alpha=3$ (dotted line). \emph{"different scales"}}
\label{FigCalphaM3}
\end{figure}
\begin{figure}[tbp]
$%
\begin{array}{cc}
\epsfxsize=6cm \epsffile{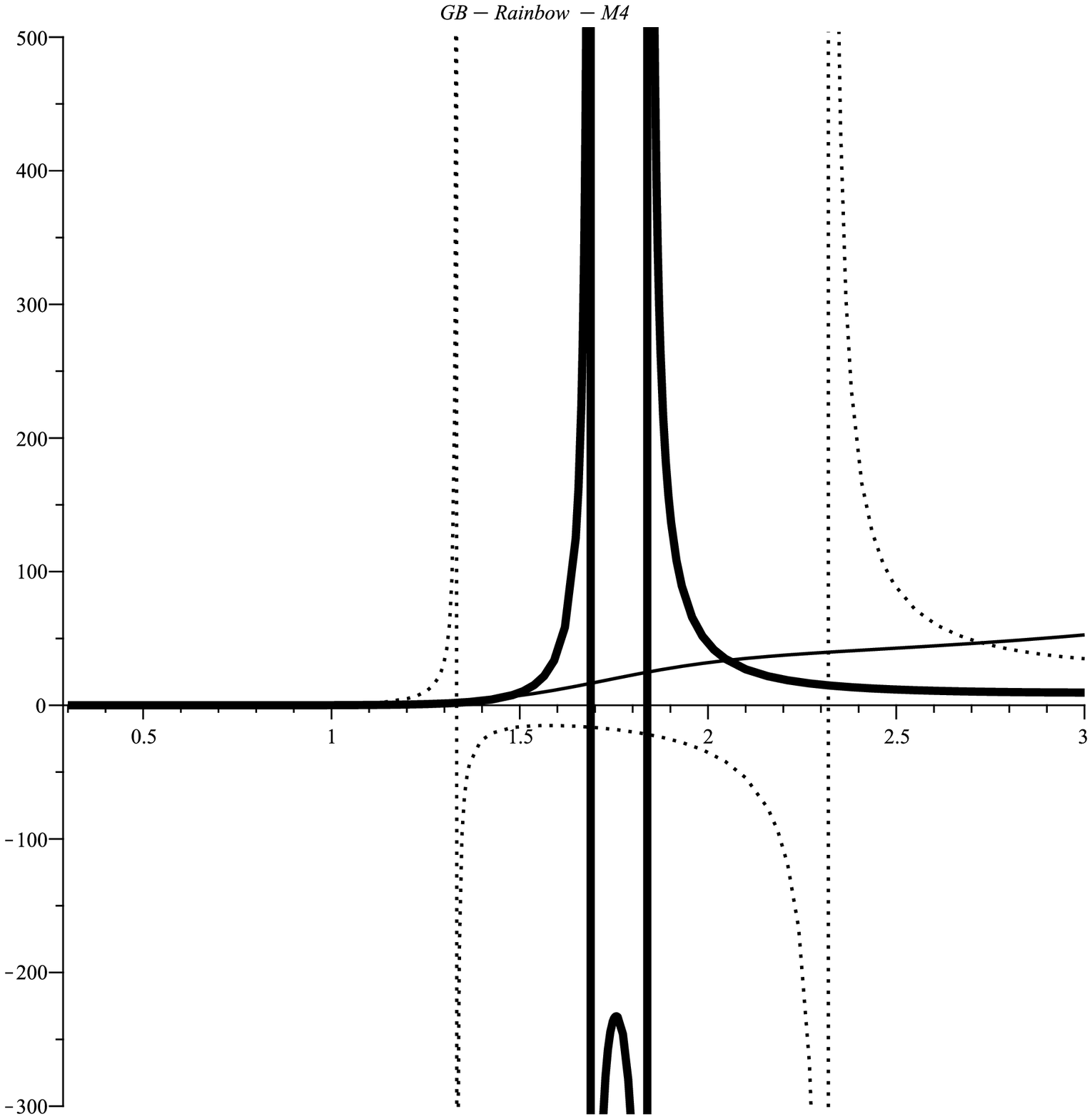} & \epsfxsize=6cm
\epsffile{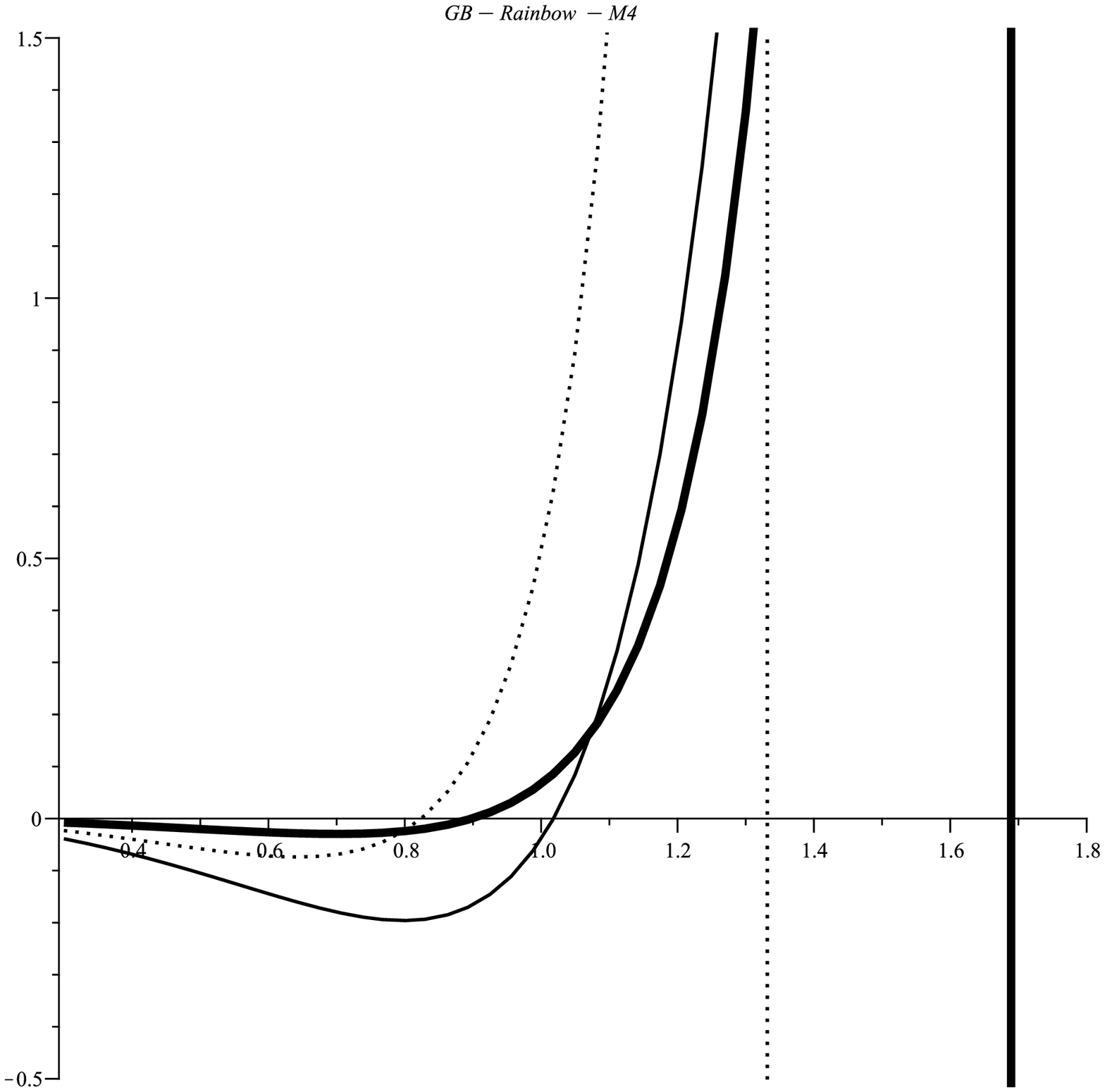}
\end{array}
$. \caption{\textbf{Model 3:} $C_{Q}$ versus $r_{+}$ for $E=1$,
$E_{p}=5$, $\Lambda =-1$, $q=1$, $\protect\lambda=2$, $d=6$, and
$\protect\alpha=0.5$ (solid line), $\protect\alpha=1$ (bold line)
and $\protect\alpha=1.5$ (dotted line). \emph{ "different
scales"}} \label{FigCalphaM4}
\end{figure}

\begin{figure}[tbp]
$%
\begin{array}{cc}
\epsfxsize=6cm \epsffile{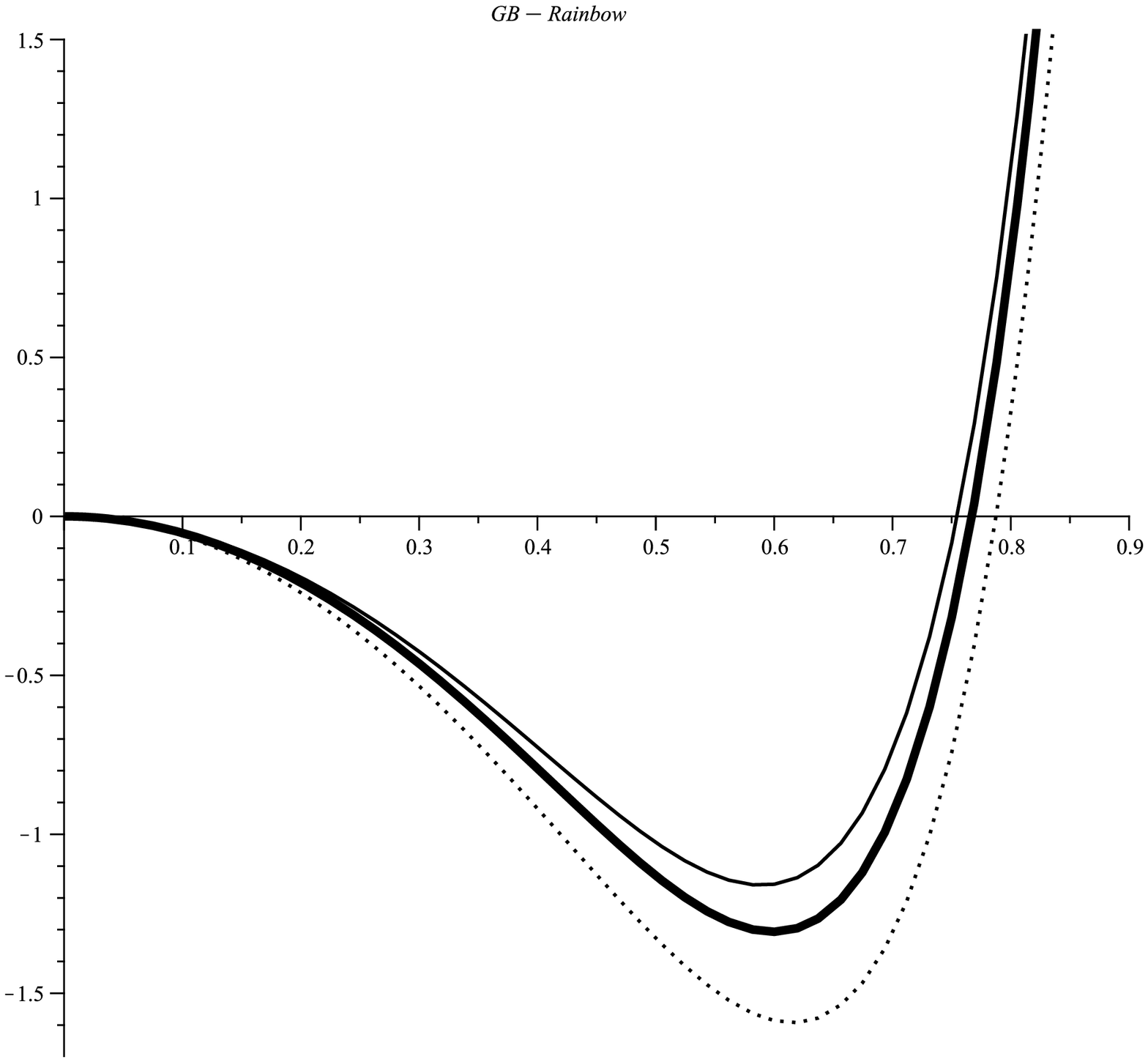} & \epsfxsize=6cm %
\epsffile{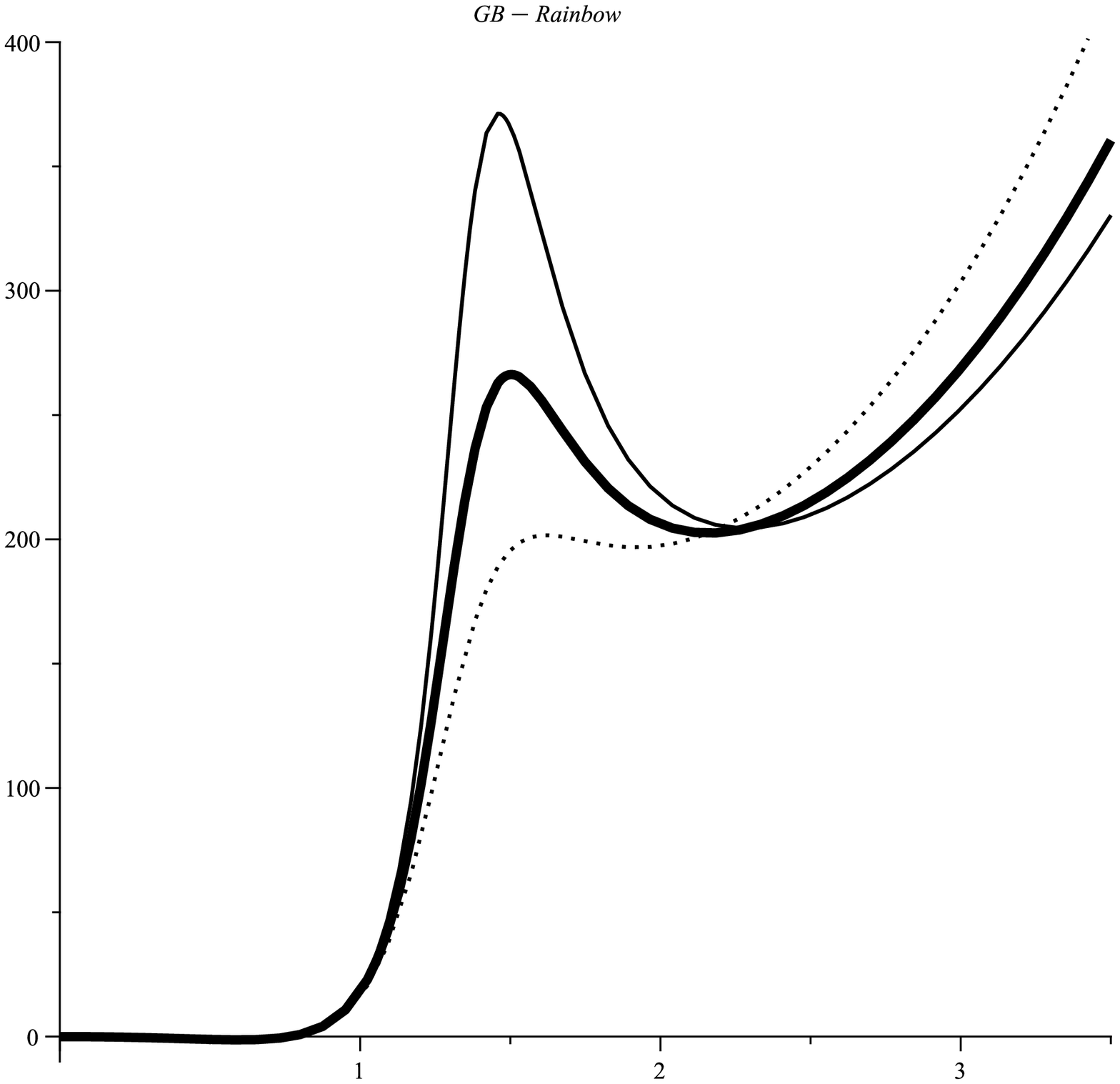}%
\end{array}
$. \caption{\textbf{Model 1:} $C_{Q}$ versus $r_{+}$ for $E=1$,
$E_{p}=5$, $\Lambda =-1$, $q=1$, $\protect\alpha=2$, $n=2$, $d=6$,
and $\protect\eta=0$ (solid line), $\protect\eta=2$ (bold line)
and $\protect\eta=5$ (dotted line). \emph{"different scales"}}
\label{FigCetaSM1}
\end{figure}
\begin{figure}[tbp]
$%
\begin{array}{cc}
\epsfxsize=6cm \epsffile{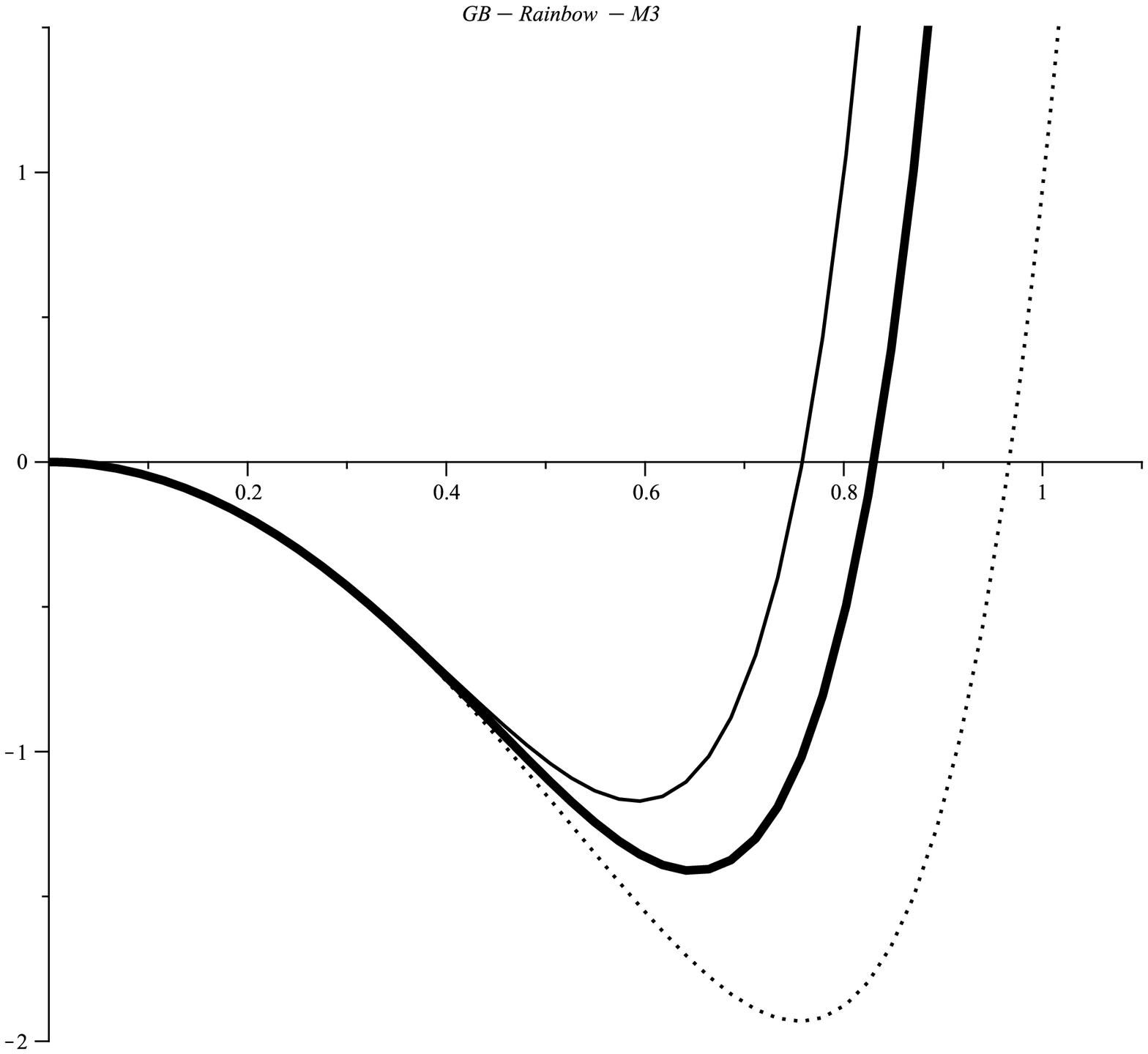} & \epsfxsize=6cm %
\epsffile{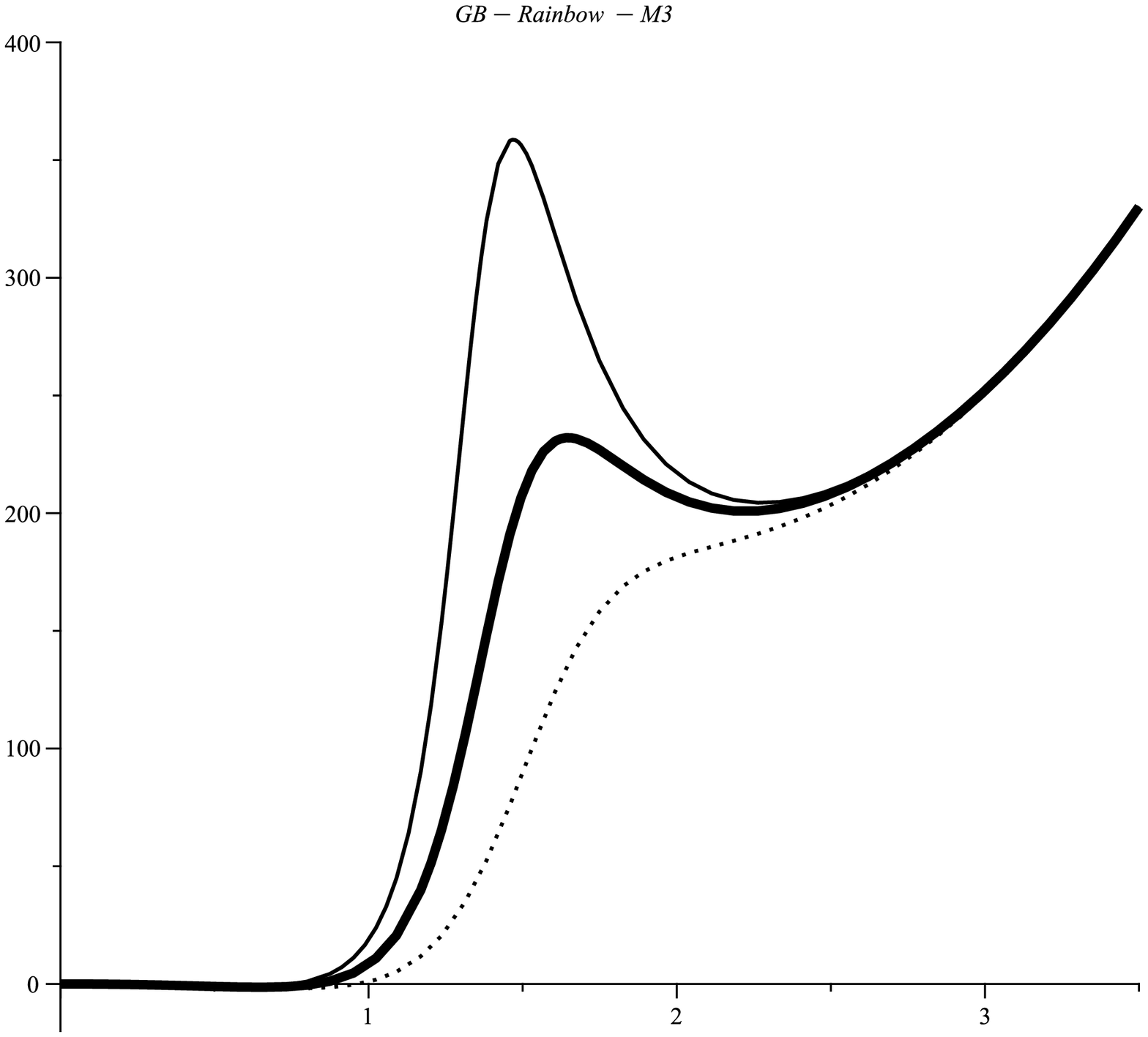}%
\end{array}
$. \caption{\textbf{Model 2:} $C_{Q}$ versus $r_{+}$ for $E=1$,
$E_{p}=5$, $\Lambda =-1$, $q=1$, $\protect\alpha=2$, $d=6$, and
$\protect\beta=0.1$ (solid line), $\protect\beta=2$ (bold line)
and $\protect\beta=5$ (dotted line). \emph{"different scales"}}
\label{FigCbetaSM3}
\end{figure}
\begin{figure}[tbp]
$%
\begin{array}{cc}
\epsfxsize=6cm \epsffile{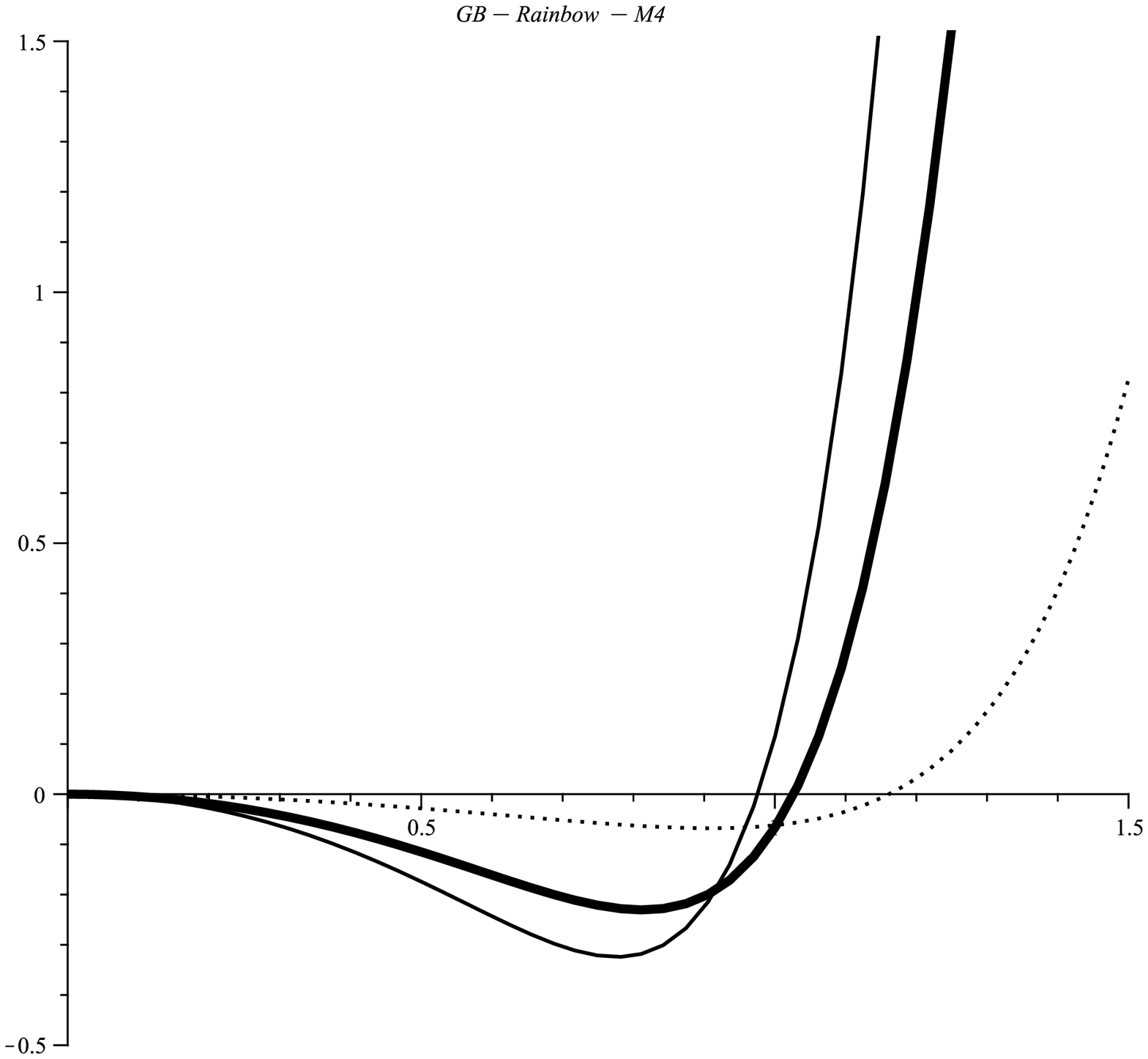} & \epsfxsize=6cm %
\epsffile{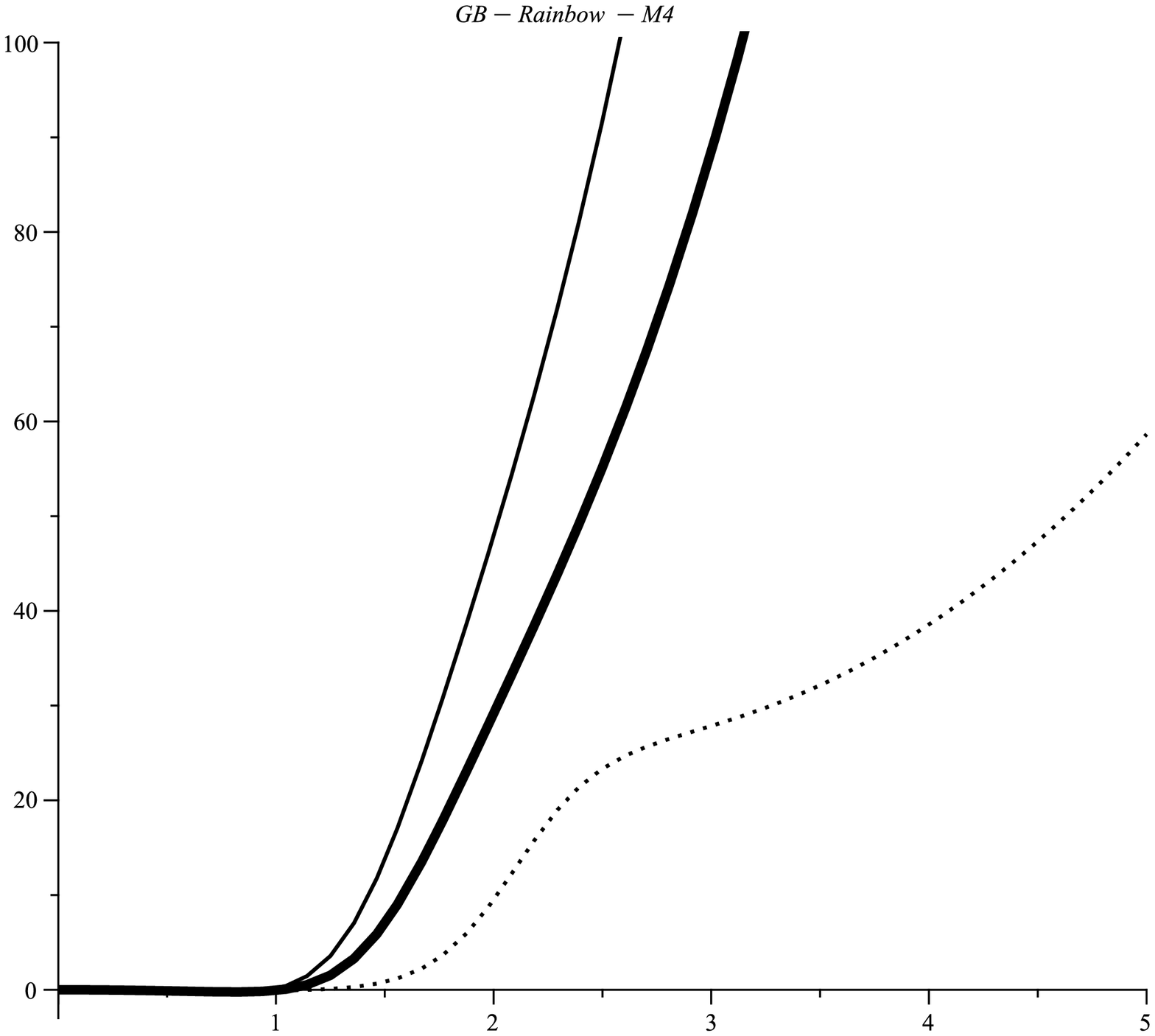}%
\end{array}
$.
\caption{\textbf{Model 3:} $C_{Q}$ versus $r_{+}$ for $E=1$, $E_{p}=5$, $\Lambda =-1$, $q=1$, $%
\protect\alpha=0.3$, $d=6$, and $\protect\lambda=0.1$ (solid line), $\protect%
\lambda=1$ (bold line) and $\protect\lambda=3$ (dotted line).
\emph{"different scales"}} \label{FigClambdaSM4}
\end{figure}

\begin{figure}[tbp]
$%
\begin{array}{cc}
\epsfxsize=6cm \epsffile{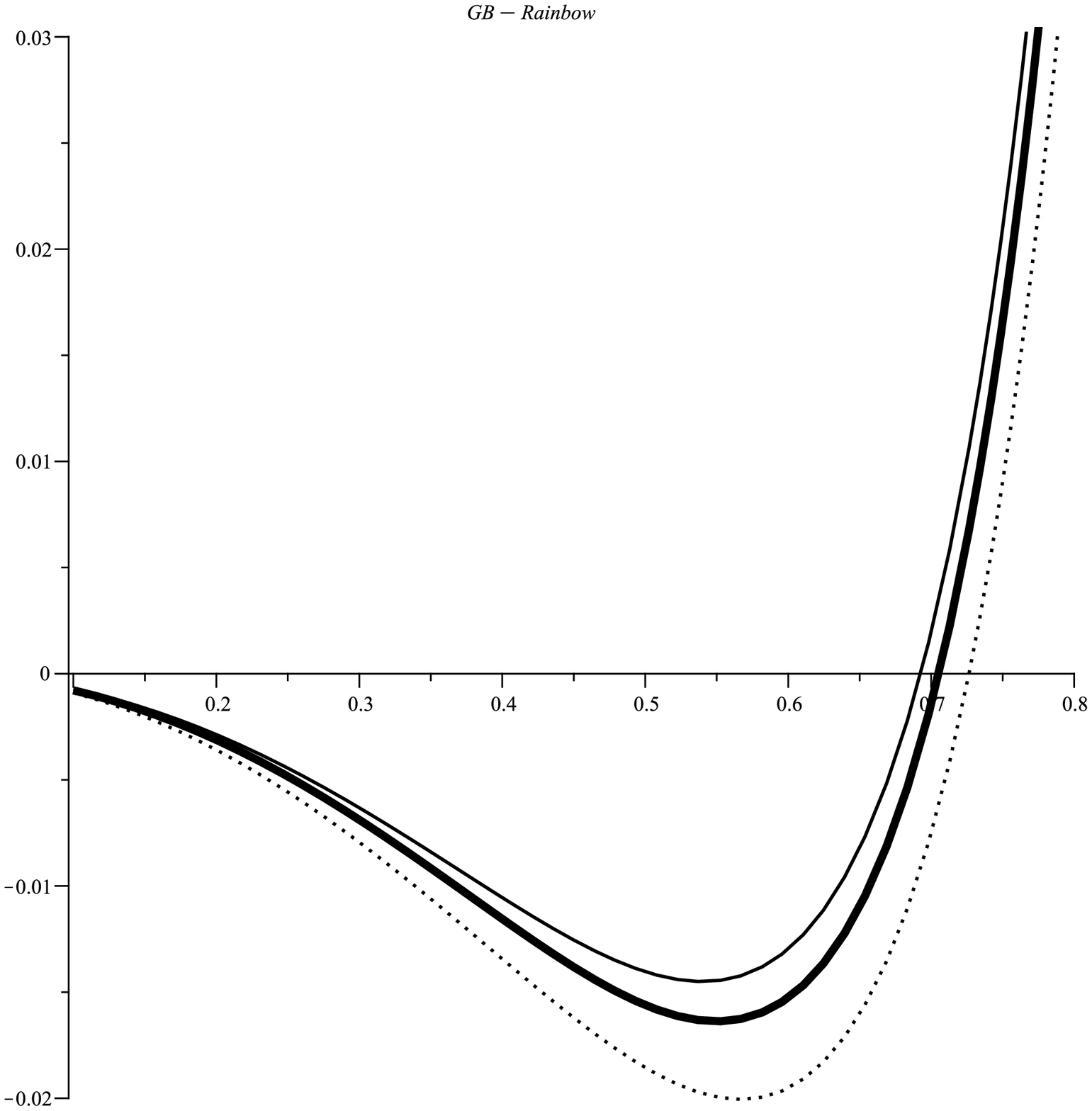} & \epsfxsize=6cm %
\epsffile{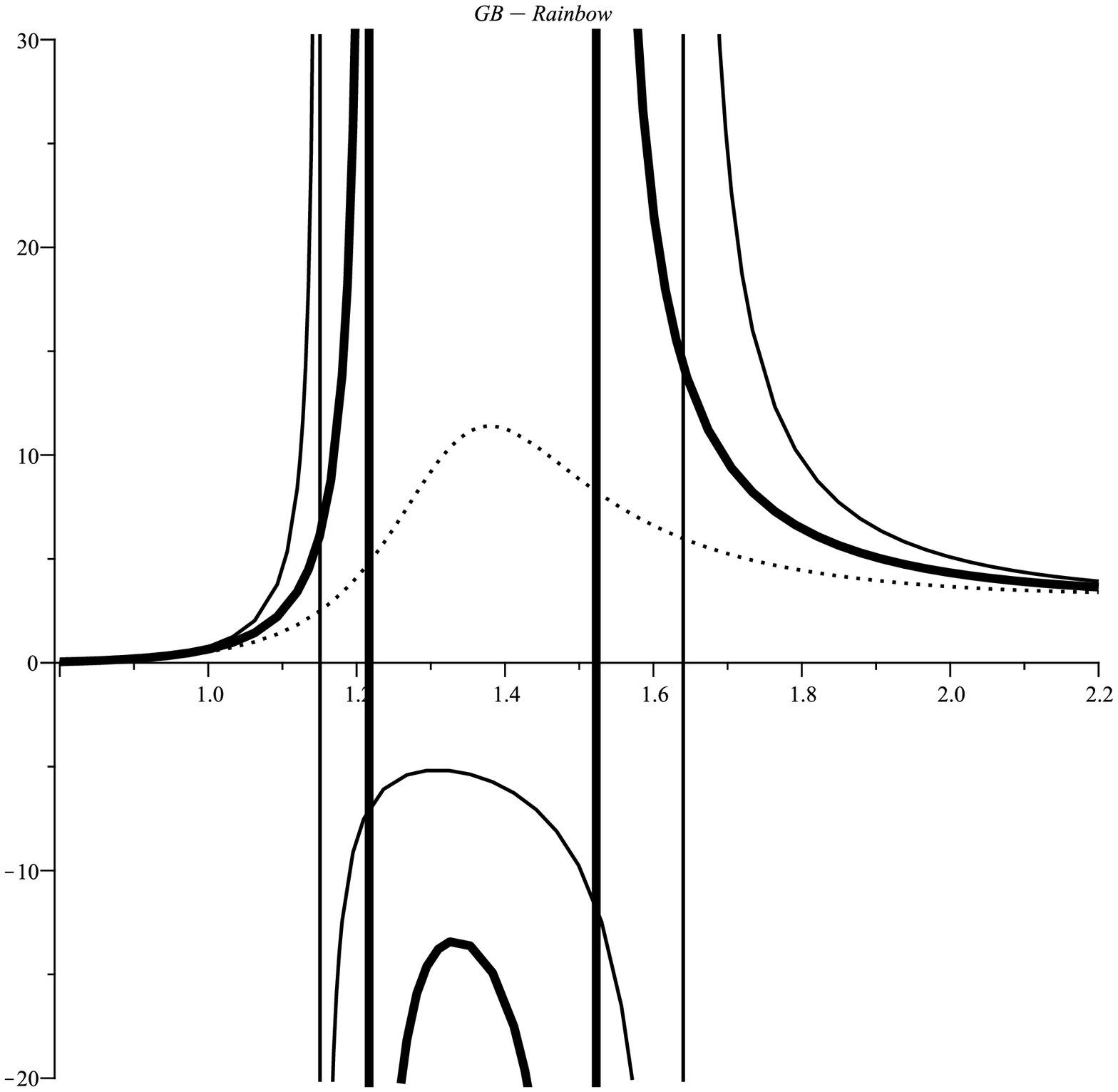}%
\end{array}
$.
\caption{\textbf{Model 1:} $C_{Q}$ versus $r_{+}$ for $E=1$, $E_{p}=5$, $\Lambda =-1$, $q=1$, $%
\protect\alpha=3$, $n=2$, $d=6$, and $\protect\eta=0$ (solid line), $\protect%
\eta=2$ (bold line) and $\protect\eta=5$ (dotted line).
\emph{"different scales"}} \label{FigCetaLM1}
\end{figure}
\begin{figure}[tbp]
$%
\begin{array}{cc}
\epsfxsize=6cm \epsffile{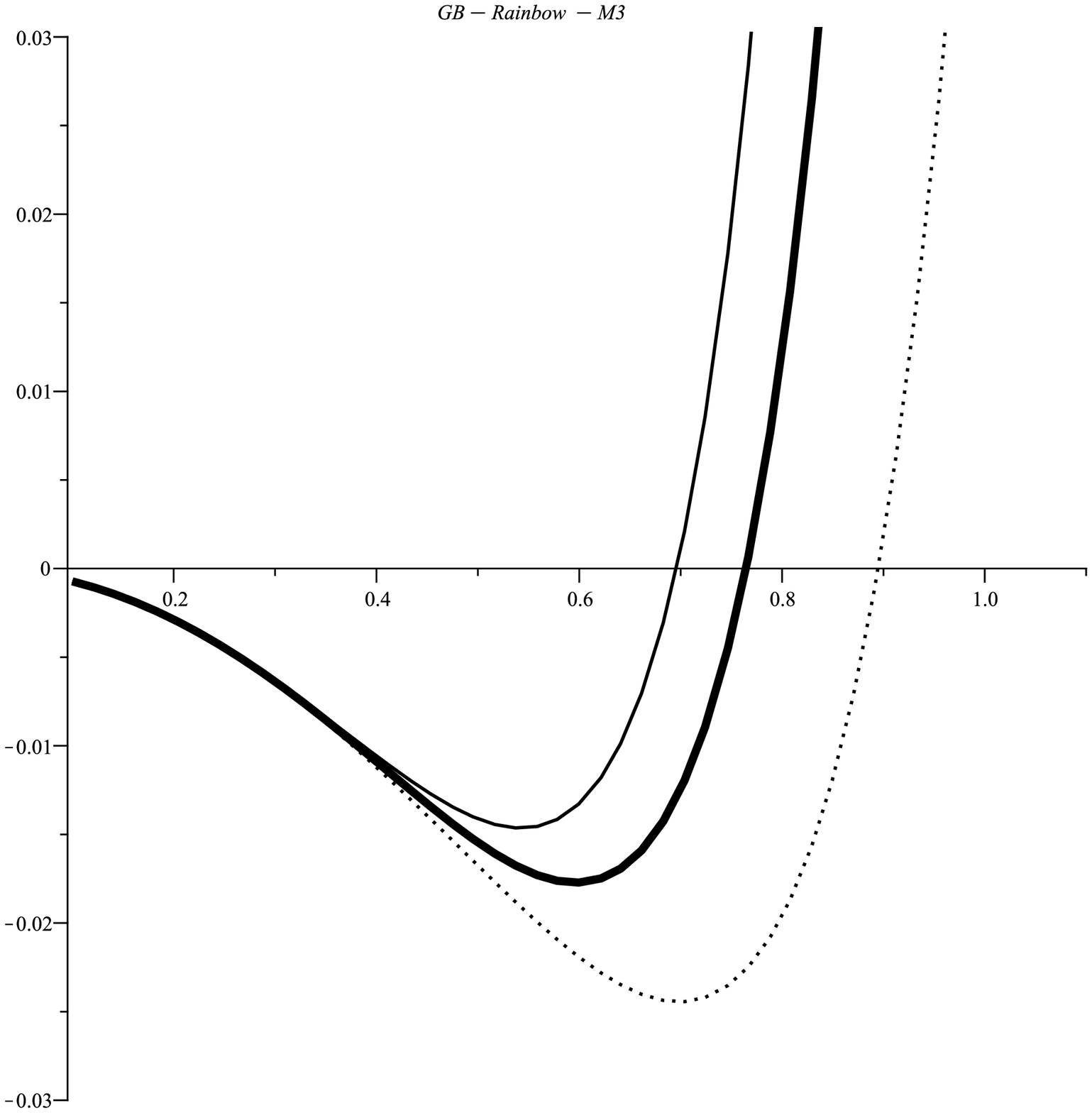} & \epsfxsize=6cm %
\epsffile{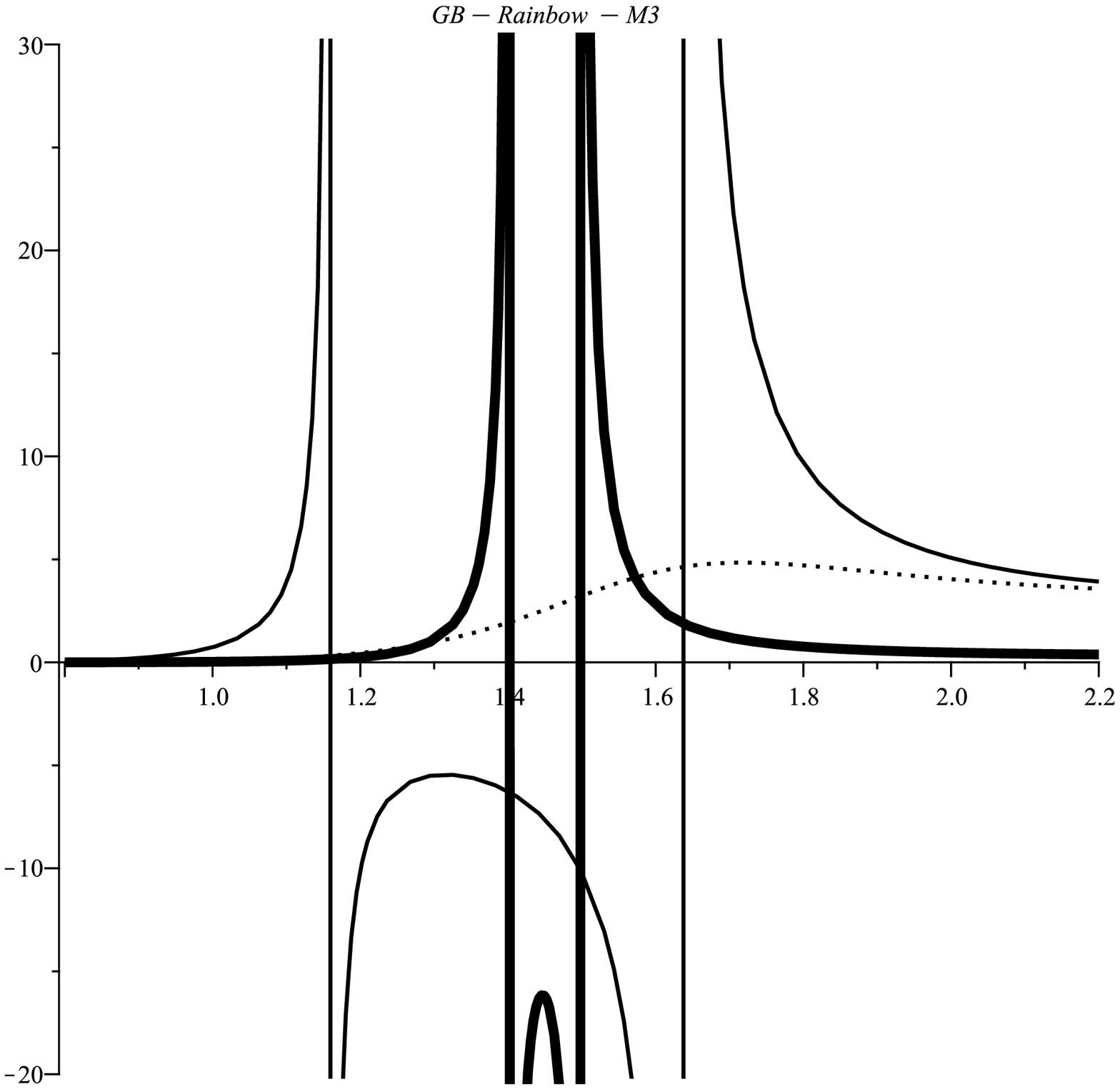}%
\end{array}
$.
\caption{\textbf{Model 2:} $C_{Q}$ versus $r_{+}$ for $E=1$, $E_{p}=5$, $\Lambda =-1$, $q=1$, $%
\protect\alpha=3$, $d=6$, and $\protect\beta=0.1$ (solid line), $\protect%
\beta=2$ (bold line) and $\protect\beta=5$ (dotted line).
\emph{"different scales"}} \label{FigCbetaLM3}
\end{figure}
\begin{figure}[tbp]
$%
\begin{array}{cc}
\epsfxsize=6cm \epsffile{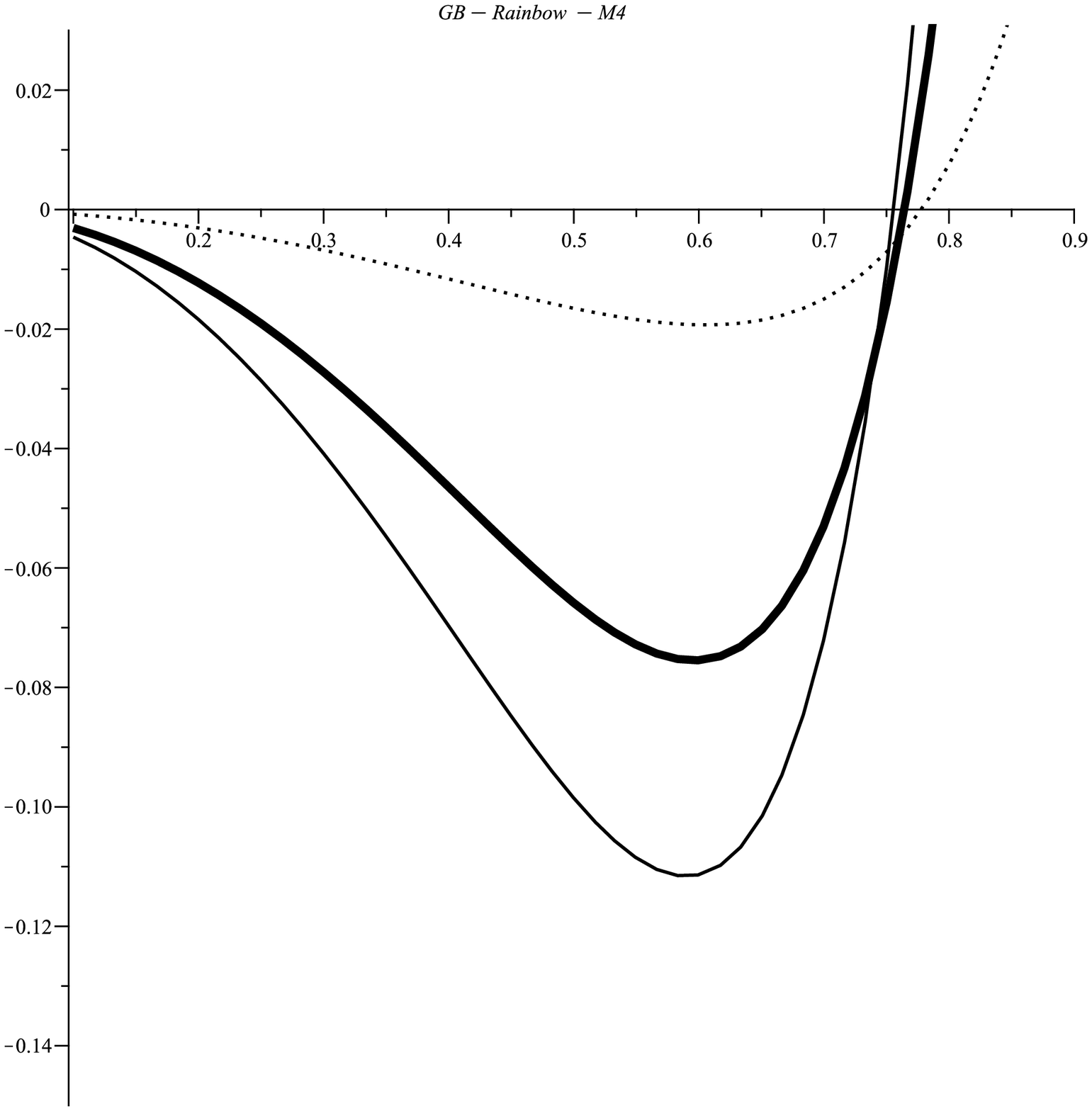} & \epsfxsize=6cm %
\epsffile{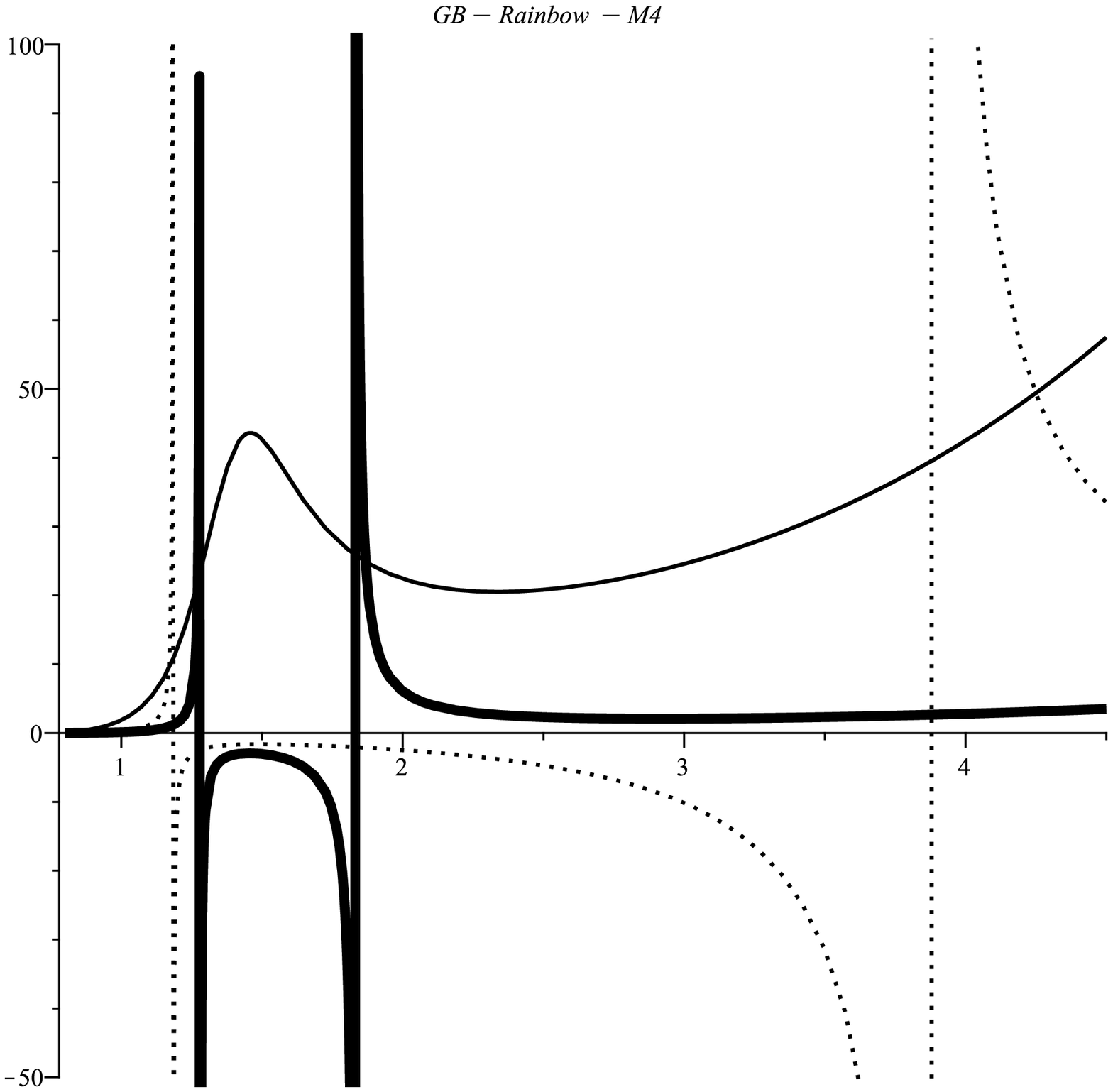}%
\end{array}
$.
\caption{\textbf{Model 3:} $C_{Q}$ versus $r_{+}$ for $E=1$, $E_{p}=5$, $\Lambda =-1$, $q=1$, $%
\protect\alpha=2$, $d=6$, and $\protect\lambda=0.1$ (solid line), $\protect%
\lambda=1$ (bold line) and $\protect\lambda=3$ (dotted line).
\emph{"different scales"}} \label{FigClambdaLM4}
\end{figure}

\begin{figure}[tbp]
$%
\begin{array}{cc}
\epsfxsize=6cm \epsffile{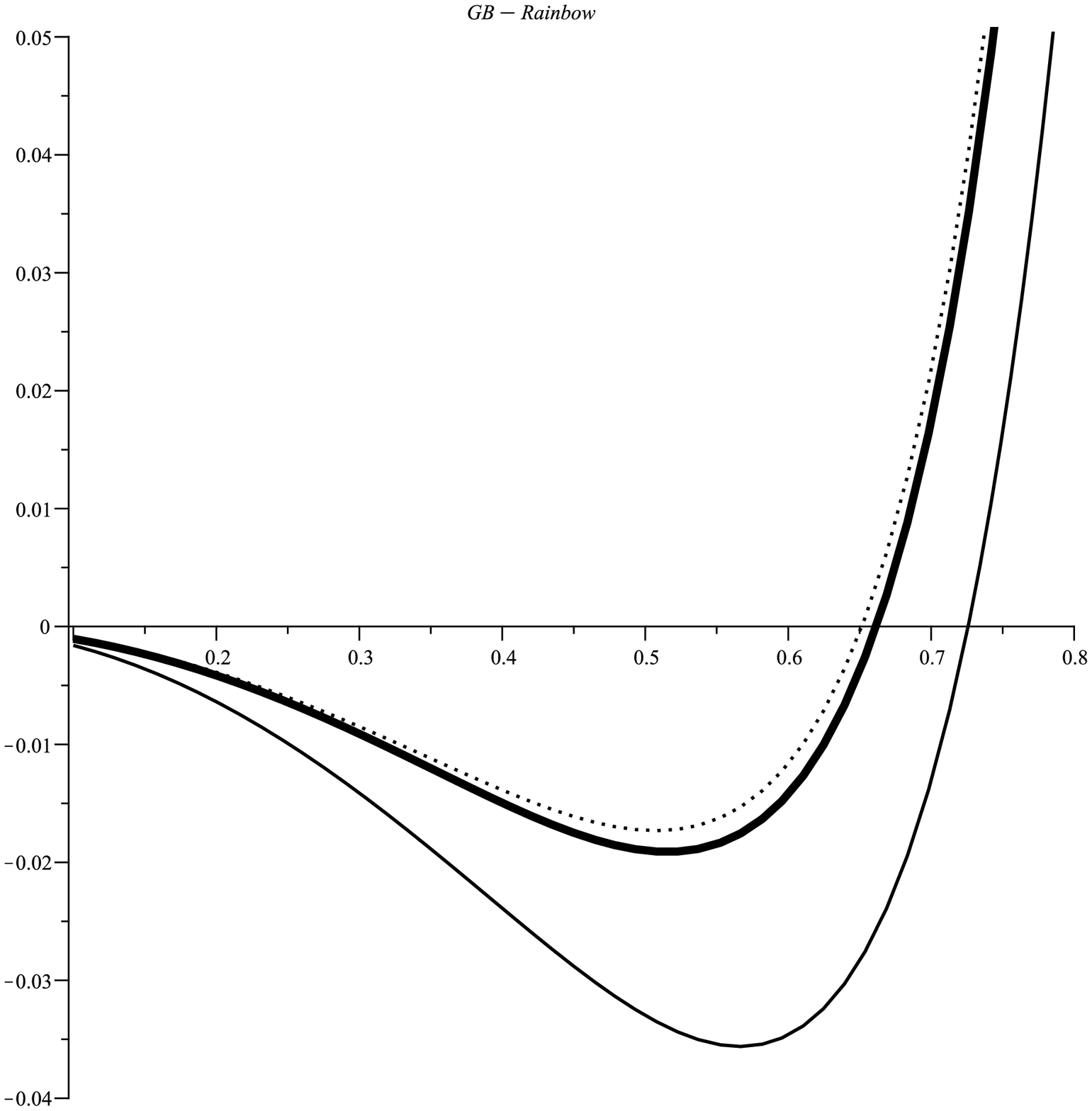} & \epsfxsize=6cm %
\epsffile{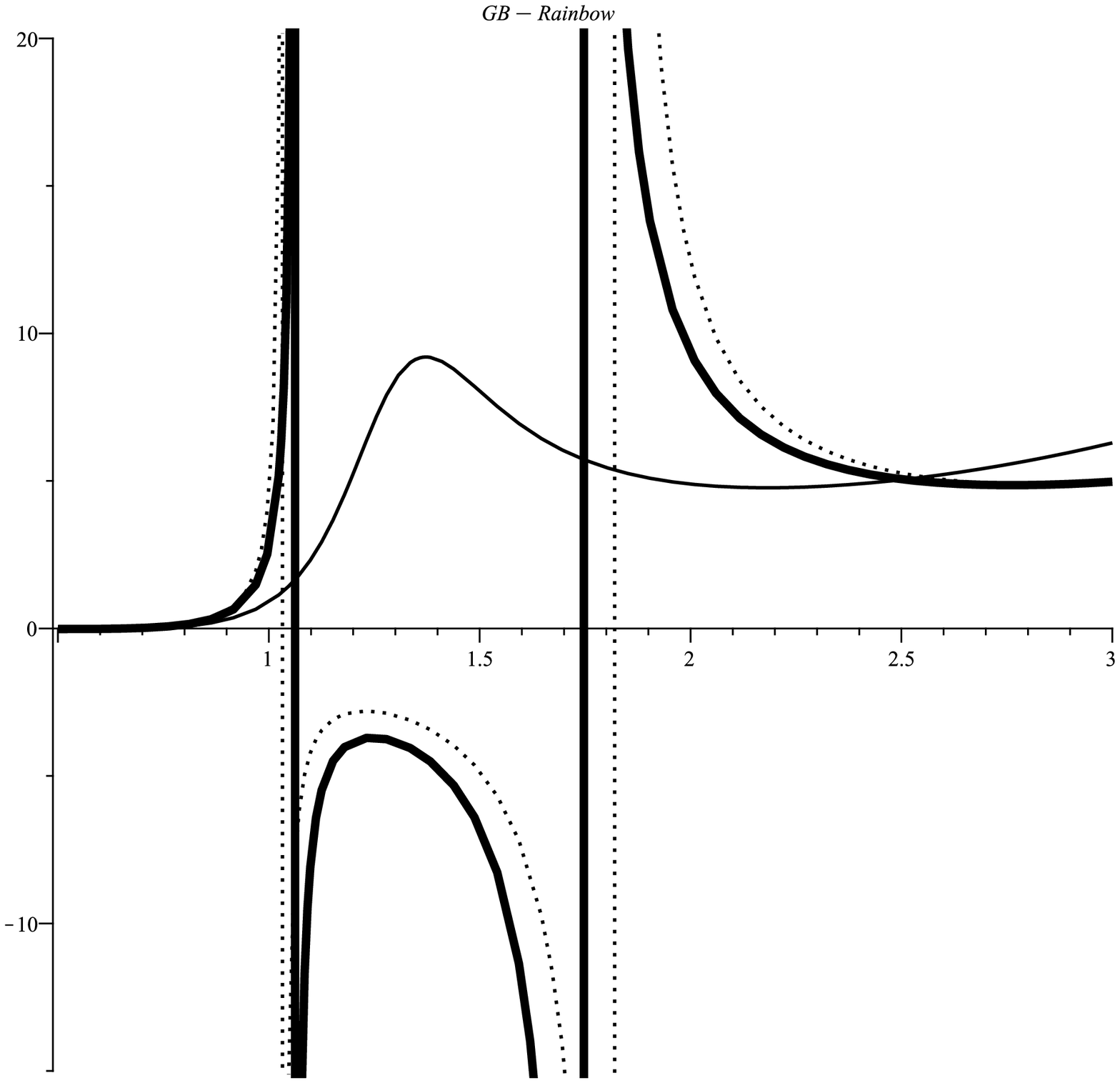}%
\end{array}
$.
\caption{\textbf{Model 1:} $C_{Q}$ versus $r_{+}$ for $E=1$, $E_{p}=5$, $\Lambda =-1$, $q=1$, $%
\protect\alpha =4$, $\protect\eta =2$, $d=6$, and $n=1$ (solid
line), $n=2$ (bold line) and $n=3$ (dotted line). \emph{"different
scales"}} \label{FigCnM1}
\end{figure}

In order to discuss the effects of gravity's rainbow on thermal
stability, we will analyze various different choices of rainbow
functions. We plot heat capacity versus horizon radius for these
different choices of rainbow functions. We also investigate the
effects of $d$ (spacetime dimensions), $\alpha $ (GB parameter)
and the rainbow parameters ($\eta, n, \beta, \lambda$ on the heat
capacity.

The head capacity also depends on the dimensions of spacetime, as
can be observed from Figs.
\ref{FigCdimensionM1}-\ref{FigCdimensionM4}. These figures
demonstrate that divergent points appear in higher dimensions.
Depending on the values of the parameters, one may find that for
low dimensional cases, $C_{Q}$ is a smooth function and with a
real positive root. However, by increasing the dimensions of
spacetime, divergences  can occur  for $C_{Q}$. In fact, two
divergences occur and $C_{Q}$ is negative in between these two
divergent points. This makes the system unstable. Although the
global behavior of $C_{Q}$ is uniform for different choices of
rainbow functions, their local properties such as slopes and the
location of the root, along with the occurrence of divergences are
different for different choices of rainbow functions.

All figures indicate that $C_{Q}$ has a real positive root at
$r_{+0}$ for all values of parameters. The location of $r_{+0}$
will be affected by variation of the parameters. It can be
observed from Figs. \ref{FigCdimensionM1}-\ref{FigCnM1}, that by
increasing $d$, $\alpha $ and $n$ the value of $r_{+0}$ decreases.
Furthermore, by increasing $\eta$, $\beta$ and $\lambda$ the value
of $r_{+0}$ increases. It is possible to choose special values of
the parameters such that the heat capacity does not diverge and
remains a smooth function. However,  it is also possible to fix
the free parameters in such a way that $C_{Q}$ has two real
positive roots at $r_{+1}$ and $r_{+2}$ ($r_{+1}<r_{+2}$). In
Figs. \ref{FigCdimensionM1}-\ref{FigCnM1}, we find that there is a
critical value for the GB parameter ($\alpha _{c}$), such that
$\alpha>\alpha _{c}$, and for this value of $\alpha$, the heat
capacity has two divergent points. Furthermore, $C_{Q}$ is a
regular function for $ \alpha <\alpha _{c}$. It may be noted that
the value of $\alpha _{c}$ depends on the values of other
parameters. In addition, considering the case of having two
divergent points in Fig. \ref{FigCetaLM1}, we find that  one can
obtain a smooth $C_{Q}$ by increasing the value of $\eta$ and
$\beta$ (decreasing $\lambda$). The same behavior can be obtained
by   decreasing the value of $n$. In other words, since we regard
$ 0<\frac{E}{E_{p}}<1$, one finds that the effects of varying
$\eta$ and $\beta$ are opposite to the effects of varying $n$ and
$\lambda$. Thus, the effect of gravity's rainbow decreases by that
decreasing $\eta, \beta$ and increasing $n, \lambda$.

\section{Conclusion}

In this paper, motivated by the developments in gravity's rainbow,
we have developed a generalization of Gauss-Bonnet gravity to an
energy dependent version of this theory. This new theory called
Gauss-Bonnet gravity's rainbow has been used  for analyzing black
hole solutions. In this theory, the metric in the Gauss-Bonnet
gravity is made energy dependent in such a way that the regular
Gauss-Bonnet gravity is recovered  in the infrared limit.  We have
explicitly calculated the modifications to the thermodynamics of
black holes in Gauss-Bonnet's gravity's rainbow. This was done by
using various phenomenologically motivated rainbow functions.  We
used the rainbow functions whose construction was motivated by
hard spectra of gamma-ray bursters at cosmological distances
\cite{1z}. We also used the rainbow functions whose construction
was motivated from results that have
been obtained in  loop quantum gravity and non-commutative geometry  \cite%
{2z}. Finally, as it is possible to construct the rainbow
functions in which the velocity of light is constant \cite{4z}, we
also used these to analyze our results corresponding to such
rainbow functions.  We have also demonstrated that  the first law
of thermodynamics still holds for this modified thermodynamics.
Finally, we commented about the thermodynamical stability of these
solutions. We showed that $C_{Q}$ has two different behaviors. In
the first case it is a smooth and regular function with a real
positive root at $r_{+0}$. In this case black hole solutions are
thermally stable for $r_{+}>r_{+0}$. For the second case, there
are two singular (divergence) points for the heat capacity in
which they are corresponding to second order phase transition
points. We showed that variation of the model parameters,
dimensionality and GB parameter can affect on the heat capacity
behavior.

It may be noted that in higher dimensions it is possible to
construct various interesting solutions to black holes  like black
rings and black saturns. The black rings \cite{2A}, and black
saturns \cite{4A} have been analyzed using  gravity's rainbow. It
has been demonstrated that in gravity's rainbow these solutions
also have a remnant. It will be interesting to construct black
rings and black saturn solutions in Gauss-Bonnet gravity's
rainbow.  It is expected that these solutions will also have a
remnant because of the modification of the thermodynamics  due to
the rainbow functions. It has also been recently argued that the
energy needed to create a mini black hole  in a particle
accelerator will increase in gravity's rainbow. Thus, it would be
possible to test a certain limit  of gravity's rainbow at the LHC
\cite{1y}. This increase in energy occurs as the energy needed to
form a mini black  hole at the LHC has to be greater than the
energy of the black hole remnant. It will be interesting to
analyze  a similar situation using Gauss-Bonnet gravity's rainbow.
It has been proposed that gravity's rainbow can be used to address
the black hole information paradox\cite{2y}-\cite{y2}. So, it
would also be interesting  to repeat such an analysis for
Gauss-Bonnet gravity's rainbow.

\begin{acknowledgements}
S. H. H. thanks the Shiraz University Research Council. The work
of S. H. H has been supported financially by the Research
Institute for Astronomy and Astrophysics of Maragha, Iran.
\end{acknowledgements}

\end{document}